\documentstyle[aasms4]{article}
\def\fdg{\hbox{$.\!\!^\circ$}}
\def\deg{\hbox{$^\circ$}}
\def\plus{{\tiny $^{+}$}}
\def\ipt{\scriptsize}
\def\foot{\footnotesize}

\begin{document}

\title{Emission-Line Galaxy Surveys as Probes of the\\
Spatial Distribution of Dwarf Galaxies.\\ 
I. The University of Michigan Survey}

\author{Janice C. Lee\altaffilmark{1} and John J. Salzer\altaffilmark{2}}
\affil{Astronomy Department, Wesleyan University, Middletown, CT 06459}
\affil{jlee@as.arizona.edu, slaz@astro.wesleyan.edu}

\author{Jessica L. Rosenberg}
\affil{Department of Physics \& Astronomy, University of Massachusetts, Amherst, MA 01003}
\affil{rosenber@umbriel.astro.umass.edu}

\author{Daniel A. Law}
\affil{Department of Engineering, University of Oxford, Oxford, OX1 3PJ, UK}
\affil{dan.law@eng.ox.ac.uk}

\altaffiltext{1}{current address: Department of Astronomy, University of Arizona, Tucson, AZ 85721}
\altaffiltext{2}{NSF Presidential Faculty Fellow} 

\begin{abstract}
Objective-prism surveys which select galaxies on the basis of line-emission are extremely effective at detecting low-luminosity galaxies and constitute some of the deepest available samples of dwarfs.  In this study, we confirm that emission-line galaxies (ELGs) in the University of Michigan (UM) objective-prism survey (\cite{mac77}-1981) are reliable tracers of large-scale structure, and utilize the depth of the samples to examine the spatial distribution of low-luminosity (M$_{B} > $ -18.0) dwarfs relative to higher luminosity giant galaxies (M$_{B} \leq$ -18.0) in the Updated Zwicky Catalogue (\cite{fal99}).   New spectroscopic data are presented for 26 UM survey objects.  We analyze the relative clustering properties of the overall starbursting ELG and normal galaxy populations, using nearest neighbor and correlation function statistics. This allows us to determine whether the activity in ELGs is primarily caused by gravitational interactions.  We conclude that galaxy-galaxy encounters are not the sole cause of activity in ELGs since ELGs tend to be more isolated and are more often found in the voids when compared to their normal galaxy counterparts.  Furthermore, statistical analyses performed on low-luminosity dwarf ELGs show that the dwarfs are less clustered when compared to their non-active giant neighbors.  The UM dwarf samples have greater percentages of nearest neighbor separations at large values and lower correlation function amplitudes relative to the UZC giant galaxy samples.  These results are consistent with the expectations of galaxy biasing.
\end{abstract}


\section{Introduction}
Studies which improve our understanding of the dependence of galaxy clustering on global galaxian properties such as morphology, luminosity and star-formation activity, are essential for deducing the correct model of galaxy formation.  There is some expectation that the underlying mass distribution is smoother than the distribution of light, since it is unlikely that formation processes have been efficient enough to produce a one-to-one correspondence between dark matter halos and visible galaxies.  Currently, the extent to which the overall galaxy distribution is biased relative to the underlying mass distribution is highly debated.  The physical mechanisms which determine the large-scale biasing relation are complex, and presumably involve an interplay of processes such as gas cooling, star formation, supernova feedback and merging (\cite{dek87}).  These processes may have affected different subsets of galaxies to varying degrees, with the result that the subsets now exhibit differing levels of large-scale clustering.  Models of galaxy formation may therefore be constrained by testing for galaxy biasing through the statistical analysis of variations in the relative spatial distribution of different types of objects.

Early work investigated the relationship between galaxy morphology and density environments (\cite{dav76,dre80}), and the observational fact that early-type galaxies are more clustered than late-type galaxies on small scales is now well-known.  Since then, the wealth of data provided by major redshift surveys over the past decade (e.g., CfA2, \cite{gel89}; SSRS2, \cite{dac94}, 1998; Stromlo-APM, \cite{lov96}; Las Campanas Redshift Survey, \cite{she96}), has allowed for the investigation of the dependence of galaxy clustering on other properties such as color (\cite{tuc95}), surface brightness (e.g., \cite{thu87}, 1991; \cite{bot86}, 1993), luminosity (e.g., \cite{ede89,sal90,lov95,wil98}) and star-formation activity (e.g., \cite{sal89,ros94,tel99}).  In this work, we examine the relative spatial distributions of sub-samples of the overall galaxy population defined by two of these parameters, star-formation activity and luminosity.

Although it has been generally well-established that star-forming galaxies are preferentially found in lower density environments than quiescent `normal' galaxies (\cite{iov88,sal89,ros94,pus94,tel95,lov99}), it is not yet clear whether a relationship between clustering and luminosity exists.  While some analyses suggest that clustering and luminosity are uncorrelated (e.g., \cite{ede89,bin90,thu91,wei91}), there is also evidence that dwarf galaxies are more weakly clustered than the population of brighter giants (\cite{sal90,san90,par94,lov95,wil98}).  Resolving this discrepancy is of particular importance in constraining galaxy formation scenarios that involve biasing (``biased galaxy formation;'' see for example \cite{dek87,whi87a,kau97}), since the majority of these models predict a rise in clustering with increasing galaxian luminosity.

Nearly all of the above studies which investigate the dependence of clustering on luminosity (except for Eder et al. (1989) and Loveday et al. (1995)) draw their samples of dwarfs from surveys which are limited in magnitude at $\sim$15.5.  Consequently, the intrinsically faint galaxies used in these analyses are confined to a small, local volume -- few dwarfs beyond the Local Supercluster are included.  This makes it difficult to draw definitive conclusions about the relative spatial distribution of dwarf and giant galaxies since the samples which are compared probe vastly different depths.  

In this study, we take a different approach to examining the clustering properties of high and low-luminosity galaxies.  We augment the depth of the typical dwarf sample by combining dwarf emission-line galaxies from the University of Michigan (UM) objective-prism survey (\cite{mac77}-1981) with dwarfs from the magnitude-limited Updated Zwicky Catalogue (UZC) (\cite{fal99}).  The spatial distribution of this population of objects is then analyzed relative to samples of giant galaxies from the UZC.  It has been shown that objective-prism surveys which select objects on the basis of line-emission are extremely effective at detecting intrinsically faint galaxies and constitute some of the deepest available samples of dwarfs (\cite{sal89}).  This ability of emission-line surveys to sample further down the luminosity function partially helps to overcome the constraints imposed by magnitude-limited galaxy samples, in which the relatively rare luminous galaxies are over-represented while the more common dwarf galaxies are grossly under-represented.  We refer to this selection effect throughout the paper as the {\it Malmquist effect} \footnote{Not to be confused with the term Malmquist bias, which has through common usage come to have a different meaning in extragalactic astronomy.} (a.k.a. the Scott (1956) effect). 

We stress that the emission-line selected dwarf samples used in this study act as surrogates for samples of the overall dwarf galaxy population.  This is out of necessity, since existing magnitude-limited redshift surveys do not detect dwarfs at distances far beyond the Local Supercluster.  For example, a dwarf galaxy with M$_{B}$ = $-$17.0 will only be included in a m$_{B}$ = 15.5 magnitude-limited survey if its velocity is under 2400 km s$^{-1}$.  For such nearby objects redshift is not an accurate measure of distance due to local velocity perturbations.  In order to study the dwarf galaxy population beyond the Local Supercluster, selection techniques that are sensitive to dwarf galaxies at greater distances must be used.  As mentioned above, surveys for emission-line objects provide samples of dwarfs which extend well beyond the Local Supercluster.  Of course, there is no guarantee that the spatial distribution of line-selected galaxies will faithfully reflect that of the overall dwarf population.  This is an important caveat which must be kept in mind as one interprets this work.  While it would be preferable to carry out this type of study using a more representative sample of dwarf galaxies (including, for example, low-surface-brightness systems), the line-selected UM catalog is one of the few available with a large sample of dwarf galaxies beyond the Local Supercluster for which redshifts exist.

Nevertheless, we do have reason to believe that our results will not be completely dominated by selection effects.  The primary concern in using ELG dwarfs as probes of the overall dwarf population is that their activity may be a consequence of their clustering environment.  {\em If} the star-formation episodes which gives rise to strong emission lines are triggered by galaxy-galaxy interactions, then line-selected dwarfs will be found, on average, in higher-density environments than quiescent dwarfs.  Thus, clustering statistics calculated for the overall dwarf population using ELG dwarf samples will be overestimated.  Since previous analyses have indicated that these strong-lined dwarfs are in fact {\em less clustered}, this effect does not seem to be present in our sample.  This is not proof that emission-line dwarfs truly reflect the spatial distribution of the overall field galaxy dwarf population, but it at least gives us some confidence that the results we obtain will not be entirely biased.

One should also keep in mind that even if these starbursting dwarfs are proper tracers of the dwarf galaxy large-scale distribution, ELG dwarfs still only represent a small fraction of the overall dwarf population.  That is, only those dwarf galaxies currently undergoing a significant star-formation episode will be found in the UM survey.  Once this episode fades in these particular galaxies, they too will become undetectable using current redshift-survey techniques.  The fraction of dwarfs currently bursting (which is related to the so called ``duty cycle") is highly uncertain, and most likely varies in a complicated way that depends on galaxian properties such as the gas mass fraction and the shape of the underlying mass distribution.  We make no attempt to account for the large additional population of dwarfs that are at the distances probed by the UM dwarfs, but are missing from the analysis because they are currently in a quiescent phase.   

In addition to investigating the dependence of clustering on luminosity, we also re-examine the relative clustering properties of emission-line galaxies (ELGs) and normal (non-active) galaxies.  By including ELGs in all five lists of the UM survey and using a substantially more complete comparison sample of normal galaxies, we improve upon the Salzer (1989) analysis of the UM galaxies, which only examined objects in the fourth and fifth lists.  Further, we extend Salzer (1989) by computing correlation functions as well as nearest neighbor statistics.

The remainder of the paper is organized as follows.  In $\S$II, we describe the surveys from which galaxy samples are drawn.  Newly acquired spectroscopic data for ELGs in the Fall UM survey area (lists I-IV) are presented in $\S$III.  A qualitative analysis of the relative spatial distributions of the UM ELGs and normal UZC galaxies is given in $\S$IV via the visual inspection of cone diagrams of each of the UM survey regions.  The clustering properties of the UM ELGs and dwarfs are quantified in $\S$V, which describes our nearest neighbor and correlation function analyses in detail, and presents the results of the calculations.  $\S$VI summarizes our results and interprets them in the context of triggering mechanisms for galaxy activity and biasing in galaxy formation models.  Throughout this paper a value of 75 km s$^{-1}$ Mpc$^{-1}$ is adopted for H$_{o}$.

\section{Galaxy Samples}

Emission-line galaxy samples are obtained from the University of Michigan (UM) objective-prism survey (\cite{mac77}-1981).  The physical properties of the samples and selection biases of the survey are discussed in detail in Salzer et al. (1989a, b) and Salzer (1989), so only a short description will be given here.

The UM survey was conducted with the 61-cm Curtis Schmidt telescope in combination with an objective prism at the Cerro Tololo Inter-American Observatory.  Objects in this survey are imaged on Kodak IIIa-J emulsion photographic plates and are selected based on the presence of [O III]$\lambda\lambda$4959,5007 and/or [O II]$\lambda\lambda$3726,3729 emission. 

The two separate regions of sky probed by the UM survey define the volumes whose galaxian spatial distributions are analyzed in this study.  The first region resides in the autumn sky and consists of survey lists I through IV of the UM catalogue.  Collectively, the 411 deg.$^{2}$ covered by these lists form an irregular strip extending from 23$^{\rm h}$12$^{\rm m}$ to 2$^{\rm h}$20$^{\rm m}$ in right ascension and from -2\fdg5 to +7\fdg0 in declination.  The second region is located in the spring sky and is covered by UM list V.  This region encompasses a rectangular 225 deg.$^{2}$ area ranging from 11$^{\rm h}$15$^{\rm m}$ to 14$^{\rm h}$15$^{\rm m}$ in right ascension and -2\fdg5 to +2\fdg5 in declination (see Figure ~\ref{fig1}). 

Redshift data for the UM ELGs are compiled from Salzer et al. (1989a) and Terlevich et al. (1991).  Also, we include previously unpublished spectroscopic data for an additional 26 UM survey objects, which are presented in the next section.  Overall, the UM spring list provides a sample of 89 ELGs and the UM fall lists provide a sample of 179 ELGs.

The comparison samples of normal (non-active) galaxies are composites of two separate galaxy catalogues.  The new Updated Zwicky Catalogue (UZC) (\cite{fal99}), which is based upon the well-known {\it Catalogue of Galaxies and Clusters of Galaxies} (CGCG) (\cite{zwi61}), provides a sample of objects with redshifts that is 98$\%$ complete to $m_{Zw} \leq $15.5 for the right ascension ranges 20$^{\rm h} \leq \alpha \leq$ 4$^{\rm h}$, 8$^{\rm h} \leq \alpha \leq$ 17$^{h}$ and declination range -2\fdg5 $\leq \delta \leq$ 50\deg.  An unpublished compilation of galaxies provided by Giovanelli and Haynes supplies objects south of the CGCG survey limit.  The latter catalogue is a compilation of galaxies from several published lists (e.g., CGCG and Uppsala General Catalogue of Galaxies (UGC, \cite{nil73}) in the north and the Morphological Catalogue of Galaxies (MGC, \cite{vor62}) and the ESO/Uppsala Survey of the ESO(B) Atlas (\cite{lau82}) in the south).   Maps of each of the UM analysis regions are shown in Figure ~\ref{fig1}, where the sky positions of the UM ELGs and normal galaxies are indicated and the UM survey boundaries are marked with dashed lines.  Note that a few UM galaxies are located outside of the nominal survey boundaries.  Presumably this is due to slight variations in telescope pointing during the course of the UM survey.  The regions covered by the objects in the comparison sample are extended relative to the UM survey regions.  The ranges of these samples are 23$^{\rm h}$0$^{\rm m} \leq \alpha \leq $ 2$^{\rm h}$36$^{\rm m}$ and -5\fdg5 $\leq \delta \leq$ 10\fdg0 for the fall UM region, and 11$^{\rm h}$0$^{\rm m} \leq \alpha \leq $14$^{\rm h}$30$^{\rm m}$ and -5\fdg5 $\leq \delta \leq $ +5\fdg5 for the spring UM region.  The comparison samples are extended to minimize errors due to boundary effects in the calculation of the nearest neighbor and correlation function statistics.  This point will be discussed further in $\S$5.2.

The apparent and absolute magnitude distributions of the ELGs and UZC comparison galaxies in both of the UM survey areas are shown in Figure ~\ref{fig2}.  Clearly, the UM survey samples further down the luminosity function than the UZC.  In fact, the ELG distributions are more heavily weighted towards the regime in which the magnitude-limited UZC becomes ineffective at detecting objects.  Whereas the median absolute magnitudes of the UZC samples are -19.52 and -19.36 for the spring and fall regions respectively, the corresponding values for the UM ELG samples are -17.74 and -17.77.  This increased depth is due to the selection of objects by line-emission.  Because the equivalent-widths of the targeted lines, [OIII]$\lambda\lambda$4959,5007 and/or [OII]$\lambda\lambda$3726,3729, generally become larger as the luminosity of the host galaxy decreases, this technique is extremely effective at detecting intrinsically faint objects and partially helps to overcome the preferential selection of luminous galaxies at large distances (the Malmquist effect) in surveys which are limited by magnitude.  The UM survey is thus an excellent source from which to draw samples of low-luminosity galaxies.  We exploit this feature of the UM sample to study the relative clustering properties of low- and high-luminosity galaxies.

\section{Spectroscopic Data}

Two major previous studies of the UM ELGs have obtained follow-up spectra for most of the 348 galaxy candidates in the survey (Salzer, MacAlpine \& Boroson 1989a, b; Terlevich et al. 1991).  As of the summer of 1993, redshifts were available, either from these sources or the literature, for all but roughly 60 of the UM galaxies.  In an effort to acquire redshifts for as many of the remaining ELGs as possible, we obtained new optical spectra for 26 UM objects in July and August, 1993.  

All observations were obtained with the KPNO 2.1-m telescope using the Goldcam spectrograph \footnote{Observations obtained at Kitt Peak National Observatory.  KPNO is operated by AURA, Inc. under contract to the National Science Foundation.} .  The data were obtained as part of an experimental queue observing program being carried out at that time by NOAO.  The detector used was a Ford 3072 $\times$ 1024 CCD with 15 micron pixels.  Due to problems with the spectrograph camera optics, only approximately 2000 pixels were in reasonable focus, limiting the spectral range to 3600 -- 6800 \AA (at 1.49 \AA/pixel).  The grating used had 500 grooves/mm and was blazed at 5500 \AA.  A slit width of 2 arcsec was used for all observations.  Reductions followed standard procedures (e.g., Salzer et al. 1995).  Of the 26 ELG candidates observed, two were found to be QSOs (UM 57 \& 230), one was a galaxy displaying only absorption lines at z = 0.102 (UM 23), five were galactic stars (UM 39, 44, 76,174, \& 292), and 18 were emission line galaxies.  

The observational results for the ELGs are summarized in Table ~\ref{tab0}.  The magnitudes listed in column 4 of the table come directly from the UM survey lists (MacAlpine et al. 1977-1981), and should only be taken as approximate.  Column 6 lists the equivalent width of [OIII] $\lambda$5007, while columns 7 -- 9 give the logarithms of the ratio of the emission-line fluxes for [OIII]/H$\beta$, [OII]/H$\beta$, and, when available, [NII]/H$\alpha$.  Column 10 lists the Balmer decrement reddening parameter c$_{H\beta}$.  Note that for several of the galaxies, the H$\alpha$ and/or [NII] lines have been redshifted out of the available spectral range.  Since c$_{H\beta}$ was not uniformly available, the line ratios in the table have not been corrected for reddening.

All 18 ELGs observed here are starburst galaxies; none are Seyferts.  With the addition of these new spectral data, only 32 of the 348 UM ELG candidates currently lack follow-up spectra.  Most of these are quite faint (m$_B >$ 18.0).

\section{Qualitative Features of the Spatial Distribution}

The sets of cone diagrams shown in Figures ~\ref{fig3} and ~\ref{fig4} schematically present the spatial distributions of the galaxies in the fall and spring survey areas, respectively.  Each set consists of two diagrams which plot (a) galaxies listed in the UZC alone and (b) galaxies in both the UM and UZC catalogues for recessional velocities between 0 and 10,000 km s$^{-1}$.  All radial velocities are corrected for the Galaxy's motion with respect to the velocity centroid of the Local Group.  In addition, galaxies which have been identified as residing in clusters or groups have had their radial velocities corrected for the velocity dispersion of the cluster.  Table ~\ref{tab1} lists all clusters/groups for which velocity dispersion corrections have been applied, along with the central coordinates, radii, cluster mean velocity, and velocity dispersion.  These were identified by visual inspection of the cone diagrams, and the values of the parameters listed have been determined from the data in our galaxy catalog.  Any group showing obvious velocity dispersion in the radial direction (``fingers of God'') is treated as a cluster.  Comparison of the groups listed in Table ~\ref{tab1} with previous studies of group/cluster identifications (e.g., Geller \& Huchra 1983) shows good correspondence for the nearer groups, but our list includes a number of more distant clusters that were not recognized in the older redshift survey data.

Cluster membership is determined on a galaxy-by-galaxy basis and depends on both spatial proximity to the group center (i.e., the galaxy lies within the cluster radius), as well as close velocity correspondence between each galaxy and the mean group velocity (observed velocity within $\pm$3$\sigma$ of the cluster mean).  Once a galaxy is identified as a cluster member, its radial position in the cone diagram is modified in the following manner.  If the observed radial velocity of the galaxy is $\pm n\sigma$ from the cluster mean, where $\sigma$ is the observed cluster velocity dispersion and n varies between 0 and 3, then the corrected velocity is $\pm nr H_{o}/3$ from the cluster mean velocity, where $r$ is the observed radius of the cluster (in Mpc) and $H_{o}$ is Hubble's constant.  {\it This effectively collapses the observed velocity ellipsoid of a cluster down to a sphere}.  The radius $r$ is determined from the data and its value is chosen to include all galaxies that fall within the $\pm$3$\sigma$ velocity ellipsoid.  

In the cone diagram of UZC galaxies in the UM fall area (Figure ~\ref{fig3}a), the most noticeable feature is the large band of galaxies that stretches laterally across the field at $\sim$5500 km s$^{-1}$. The band begins as a narrow strip at the western edge of the plot and  widens as it extends toward the east.  This structure is part of the Pisces-Perseus Supercluster (\cite{hay86}), the large sheet-like ensemble of galaxies which extends across nearly half of the South Galactic Cap.  Five clusters can be found within the structure of the band, the richest of which is Abell 194 at $\sim$1\fh2.  Toward the eastern edge of the plot at $\sim$2$^{\rm h}$15$^{\rm m}$, the broad end of the band forms the boundary of a void.  In cone diagrams which cover a more extensive volume (\cite{fal99}), this void (diameter $\sim$1200 km s$^{-1}$) can be seen to persist northward to a declination of $\sim$ 24 degrees.

Two other voids of comparable extent inhabit the foreground.  One underpopulated region, roughly centered with respect to the right ascension of the plot, is evident from 2500 km s$^{-1}$ to 4000 km s$^{-1}$ and extends to $\delta$ $\sim$48 degrees. One UZC galaxy, 00422+0453 (M$_{B}$ = -17.8) appears in the center of this void. Immediately behind this region in the western portion of the diagram lies the second void, which begins at 4000 km s$^{-1}$ and extends to the anterior edge of the large band.  This void extends northward to a declination of at least 24 degrees.

Behind the band, a larger underpopulated volume is evident from 6000 km s$^{-1}$ to 8000 km s$^{-1}$ and from 0$^{h}$ to 2$^{h}$ in right ascension.  Five UZC galaxies, three of which are also detected by the UM survey, appear in this region.  These galaxies are 00203+0633, 00279+0536, I0073 (UM 84), 01090+0104 (UM 307), and 01189+0322 (UM 96) (M$_{B}$ = -19.7, -19.7, -19.2, -20.4, -19.3 respectively).  The cone diagram which plots the UM ELGs and UZC galaxies together (Figure ~\ref{fig3}c) reveals that there are an additional 7 ELGs (UM 47, 65, 98, 99, 119, 215, 92; M$_{B}$ = -18.7, -18.1, -18.4, -17.9, -17.1, -17.3, -16.3)  which reside in this low-density area.  These galaxies appear to be creating substructure, carving the low-density volume into smaller voids.

A comparison of Figures ~\ref{fig3}a and ~\ref{fig3}b show that the UM ELGs are concentrated in the areas where the UZC galaxies have formed structures - that is, the ELGs generally describe the same large-scale features as galaxies in the UZC.  Clearly however, the ELGs are considerably more dispersed and avoid regions of high galaxian density.  No UM ELGs appear in the clusters defined by the UZC galaxies.  This latter observation is consistent with the findings of Dressler et al. (1985), Teague (1988) and Salzer (1989), and is likely due to gas-stripping mechanisms in high density clusters which inhibit activity.  

Turning now to the cone diagram of the UZC galaxies in the UM spring region, (Figure ~\ref{fig4}a), the Local Supercluster (LS) is seen to dominate the region from 0 - 2000 km s$^{-1}$.  A filamentary structure extends outward from the LS to $\sim$3000 km s$^{-1}$ to form the boundary between two voids, one region in the east stretching from 1000 - 3000 km s$^{-1}$ which persists northward to $\sim$36 degrees, and another in the west from 2500 to 3500 km s$^{-1}$.  Three UZC galaxies, 11429+017, 11494-0222 and 12356+0014 (M$_{B}$ = -18.2, -18.1, -18.5 respectively), reside in this latter void.  Figure ~\ref{fig4}b, where the UM ELGs and UZC galaxies are plotted together, reveals that there are an additional three ELGs, UM 454, UM 455 and UM 513 (M$_{B}$ = -16.9, -16.3, -16.2 respectively) in this void.

A honeycomb-like structure appears in the region from 5000 - 7500 km s$^{-1}$.  The cluster MKW4 (\cite{mor75}) at $\sim$6000 km s$^{-1}$ extrudes from the eastern edge of the central ring of the honeycomb.  Filamentary structures extend outward in all directions from this ring, connect with other filaments and define the neighboring voids.  One UM ELG, UM 507 (M$_{B}$ = -16.2) appears within the void defined by the central ring of the honeycomb.

Again, a comparison of Figures ~\ref{fig4}a and ~\ref{fig4}b show that ELGs are more dispersed and avoid the major cluster (MKW 4) in this field.  Other clusters identified in this field are listed in Table ~\ref{tab1}.  There are a number of ELGs that reside in the neighborhood of the Local Supercluster.  Nevertheless, these ELGs do not penetrate the densest regions of the LS -- they only populate the periphery.  The reader is referred to Salzer (1989) for an in-depth discussion of the ELG spatial distribution in this field.

In summary, the qualitative analysis performed by visual inspection of these cone diagrams suggests that although the ELGs and normal galaxies outline the same large-scale features, the ELGs are less tightly confined to the features and tend to prefer regions of lower density.  That is, compared to the population of normal galaxies, the ELGs appear to be less clustered.  This suggests that a sub-sample of low-luminosity galaxies drawn from the UM survey would also be less clustered when compared to higher luminosity galaxies in the UZC. The statistical analyses conducted in the next section will quantify these results.

In Figure ~\ref{fig5} both of the UM fields have been plotted in a recessional velocity range extended to 20,000 km s$^{-1}$.  Examination of these diagrams shows a large difference in the depths of the samples.  Figure ~\ref{fig5}a shows that the normal galaxies in the UM fall region are well-represented out to a velocity of $\sim$12,000 km s$^{-1}$.  At this velocity, structures are still relatively well-defined, and a ridge of galaxies can be seen to cross the field, enclosing two voids in the foreground from 9000 to 12000 km s$^{-1}$.  Beyond 12,000 km s$^{-1}$ the normal galaxies become poorly sampled.  Inspection of the normal galaxies in the UM spring region (Figure ~\ref{fig5}b), shows that this decline occurs at $\sim$10,000 km s$^{-1}$.  Beyond the outer boundary of the well-delineated honeycomb structure centered at $\sim$8000 km s$^{-1}$, the normal galaxies become poorly sampled. 

The ELGs also suffer a decline in density with increasing distance.  The rate of decline, however, is not as severe.  A simple number count of galaxies in the fall region within the velocity range from 16,000 to 20,000 km s$^{-1}$ clearly illustrates this point -- whereas, 10.2$\%$ (18/176) of the ELGs reside in this region, a mere 1.4$\%$ (7/495) of the normal galaxy sample can be found at these distances.  In the spring UM region, the same effect is seen although there are fewer ELGs in this volume:  5.6$\%$ (5/89) of the ELG sample exists beyond 16,000 km s$^{-1}$ compared to only 0.3$\%$ (1/375) of the normal galaxy sample.  These statistics attest to the power of using ELGs to chart large-scale structure in redshift regimes where magnitude-limited surveys are ineffective at detecting objects.  Moreover, it is clear that it is not necessary to map every galaxy to gauge the general profile of the large-scale distribution.  This combination of depth and efficiency indicates that there is great potential in using ELG samples and similar ``sparse sampling'' techniques to trace galaxian structure at ever larger distances.   

\section{Statistics of the Galaxy Distribution}

\subsection{Sample Construction}

The manner in which galaxy samples are defined is of particular importance to this analysis.  Calculations are done on four groups of samples to deconvolve effects caused by the selection of ELGs as probes of the galaxian spatial distribution.  This section highlights the important qualities of the various samples and attempts to provide a framework which the reader may find useful to refer to in later discussions of the statistical analyses (see in particular, Table ~\ref{tab2}).

For each of the two regions considered in this study, three samples are constructed.  The ELGs taken from the UM survey comprise one sample.  UZC galaxies which reside within the areas covered by the UM survey form a sample of ``normal'' galaxies.  Finally, UZC galaxies and galaxies from the unpublished catalogue provided by Haynes and Giovanelli in the extended regions (defined in $\S$2) are taken as a computational comparison sample.  The need for this latter sample will be made clear in $\S$5.2.  UM ELGs which appear in the normal galaxy and computational comparison samples are deleted from them and appear in the ELG sample only.  Further, we exclude from the ELG sample a handful of objects in which the emission identified in the UM survey is due to normal disk HII regions (e.g., UM 505 and 506 from list V).

Galaxies outside the velocity range from 2000 km s$^{-1}$ to 8000 km s$^{-1}$ are removed from the ELG and normal galaxy samples for the statistical analyses carried out below.  The lower bound is necessary because radial velocity is not a reliable indicator of distance for near-by objects since peculiar velocities can be comparable to recessional velocities in the local neighborhood.  Also, boundary effects on the computation of clustering statistics can be severe at the apex of the cone diagram.  The upper bound is required to restrict the volume to well-sampled regions.  Its determination is based on the discussion of sample depth in the previous section.  As noted there, a rapid decline in the number of UZC galaxies occurs beyond a velocity of $~\sim$10,000 km s$^{-1}$ in the fall region, and beyond $~\sim$8000 km s$^{-1}$ in the spring region, while the UM ELGs are well-sampled beyond these velocities.  Therefore, a conservative upper velocity limit of 8000 km s$^{-1}$  for the UM and UZC samples is imposed to avoid the ``sampling mismatch'' that can potentially produce false results in the following statistical analyses.  At this velocity, an apparent magnitude limit of m$_{Zw}$= 15.5 corresponds to a complete sampling of normal galaxies brighter than -19.6 (using H$_{o}$ = 75 km s$^{-1}$ Mpc$^{-1}$).  This upper bound of 8000 km s$^{-1}$ thus represents a compromise between the need to retain the lower luminosity galaxies and the desire to sample the largest volume possible.  Since the computational comparison sample is meant to extend the limits of the normal galaxy sample to avoid boundary effects, no velocity restrictions have been imposed on its members.

These three samples, the ELG and normal galaxies from 2000 km s$^{-1}$ to 8000 km s$^{-1}$ and the computational comparison sample in its entirety, represent one group of data with which the nearest neighbor and correlation function statistics are calculated.  It is also necessary to perform these analyses on samples which consist of only field galaxies - galaxies which do not reside in gravitationally-bound clusters.  This is to ensure that any statistical differences found to exist between the ELG and normal galaxy spatial distributions are not due to the paucity of ELGs in dense, clustered regions (as discussed in $\S$4).  Removing galaxies which reside in clusters from consideration also improves the consistency of the morphological makeup between the samples and decreases the likelihood that differences in clustering strengths are due to the morphology-density relation (\cite{dav76,dre80}).  This relationship suggests that elliptical galaxies preferentially reside in regions of high local galaxy density while spiral galaxies favor regions of low galaxy density.  In particular, ellipticals appear to be largely confined to the virialized cores of clusters.  Since the ELG samples are dominated by later-type galaxies (spirals and irregulars), removing cluster members from both the ELG and normal galaxy samples helps ensure that all samples have similar morphological compositions.  For these reasons, a second group of data, in which galaxies residing in clusters (specified in Table ~\ref{tab1}) are eliminated from each of the samples, is constructed and analyzed.

To investigate the spatial distribution of low-mass galaxies, and test the biased galaxy formation model prediction that intrinsically faint, dwarf galaxies are less clustered than the larger more luminous galaxies, a third group of data is constructed in which a sub-sample of low-luminosity galaxies is drawn from the dwarf-rich UM ELG sample.  Objects from {\em both} the ELG and normal galaxy samples which are fainter than M$_{B}$ = -18.0 are taken as a sample of dwarfs, while UZC galaxies that are more luminous than -18.0 are taken as a sample of giants.  The division between low- and high-luminosity galaxies is made by considering the desirability of including only the most extreme dwarfs in the analysis and the competing need to include a large enough number of objects in this sample from which reliable statistics may be calculated.  As before, these samples are restricted to a velocity range of 2000 km s$^{-1}$ to 8000 km s$^{-1}$, and a computational comparison sample without velocity limits is constructed from giant normal galaxies in the extended region.  Since it is necessary to deconvolve morphology-density effects in this analysis as well, a fourth group of data from which cluster galaxies are removed from the dwarf, giant and computational comparison samples, is constructed and examined.

It should be noted that in the current analysis luminosity is used as a surrogate for mass.  Although it would be preferable to separate our dwarf and giant samples by mass, this is impractical given the available data.  While we do not believe this choice compromises our conclusions, the existence of low-surface brightness galaxies like Malin 1 (Bothun et al. 1987) with large masses and relatively low luminosities suggests that some caution should be exercised when considering these results.  We point out, however, that since our dwarf sample consists primarily of low-luminosity bursting ELGs ($\sim$86\% of the fall and $\sim$76\% of the spring dwarf samples are composed of ELGs), the dwarfs which are considered in this analysis have inflated luminosities and reduced mass-to-light ratios.  Thus, selecting by luminosity has the effect of {\it enhancing} the mass differences between dwarfs and giants, rather than blurring them.

In the following analyses, statistical quantities are calculated for a ``test'' population relative to a ``control'' population and for a control population relative to itself.  Table ~\ref{tab2} provides a summary of the contents in the four groups of data created for each region and indicates the samples which are considered test populations and those which are considered control populations.

\subsection{Nearest Neighbor Distributions}

Given a sample of objects, the distribution of nearest neighbor separations may be directly calculated by measuring the distances between a target object and all other objects in the sample, retaining the minimum value, and repeating this procedure for every object in the sample.  One can imagine that objects near the sample boundaries may have nearest neighbors which reside outside of the sample itself, so that distributions which are entirely constructed with separations found from within the sample may be skewed to larger separations.  This is the reason that the extended computational comparison sample - which is simply a sample of the control population over a larger area - is introduced.  {\em In all of the cases below, the nearest neighbors of galaxies in the sample being considered are found in the extended computational comparison sample.}  This allows the periphery of the sample volume to be searched for nearest neighbors and prevents sample boundary effects from introducing errors into the analysis.

For each group of data specified in Table ~\ref{tab2}, three different distributions of nearest neighbor separations are determined.  The first distribution compiles ``test object-control object'' nearest neighbor distances and is the analog of the cross-correlation function computed in the next section.  The second distribution compiles ``control object-control object'' nearest neighbor distances within the test volume and is the analog of the auto-correlation function, also computed in the next section.  Comparison of these two distributions provides one measure of the {\em difference} in spatial distribution between the two samples.  In particular, we apply Kolmogorov-Smirnov (KS) tests to the cumulative nearest neighbor distributions to determine the probability that the test galaxies and control galaxies are drawn from the same population of nearest neighbor separations.  The third distribution provides a baseline for comparison and represents ``random object-control object'' nearest neighbor separations within the test volume.  Calculations for 1000 simulated random samples of {\em n} galaxies, where {\em n} is equal to the number of objects in the test sample, are performed and averaged.  The velocity distributions that are used to generate the random samples are determined by assuming a magnitude-limited (m$_{Zw} \le$  15.5) sample and a luminosity distribution consistent with the normal galaxies.  For the latter, we utilize Schechter function parameters (Schechter 1976) derived from the normal galaxies in the computational comparison sample via the Lynden-Bell (1971) method.  The calculated luminosity functions and Schechter function fits are plotted in Figures ~\ref{fig6}a (fall region) and 6b (spring region).  Values for M$^{\ast}$, $\alpha$ and $\phi^{\ast}$ are indicated in the plots.

Results from the nearest neighbor calculations are presented in Figures ~\ref{fig7} (nearest neighbor separations for the ELG-normal galaxy data sets) and ~\ref{fig8} (nearest neighbor separations for the dwarf-giant galaxy data sets).  In all figures, histograms of the control and random samples have been scaled to the area under the test sample histogram.  Plots for the samples in which all galaxies are included are shown in the left column of figures, while those for the samples from which cluster members are removed are shown on the right.  The lower half of each figure shows the cumulative fraction distributions for the nearest neighbor distances plotted in the upper frame.

Examination of Figure ~\ref{fig7}a, which presents the results for the fall ELG/normal galaxy analysis where cluster members have been included, reveals that the ELGs are more dispersed than the normal galaxies.  The normal galaxy histogram exhibits a large peak at small separations (between 0 and 1 Mpc) and a rapid decline in the number of objects beyond 2 Mpc.  Very few objects are separated by larger distances.  Only 9.1$\%$ (29/318) of the normal galaxies are seen to have nearest neighbor separations that are greater than 3 Mpc.  In contrast, the ELG distribution has a much broader peak (between 1 Mpc and 3 Mpc), and 25.6$\%$ (23/90) of the ELGs have normal galaxy nearest neighbors which are greater than 3 Mpc away.  The weaker clustering of the ELGs is perhaps most evident in the cumulative histogram diagram.  The ELG curve rises more gradually than the normal galaxy curve, indicating that the ELGs are more dispersed.  Nevertheless, it is also clear that the ELGs are more strongly clustered than a random distribution of galaxies.  A KS test applied to the ELG and normal galaxy distribution shows that the hypothesis that the two samples have been drawn from the same population of nearest neighbor separations may be rejected at the 99.99$\%$ confidence level.

Removing the cluster members from the samples has a clear impact on the nearest neighbor distribution of the normal galaxies, but minimally changes that of the ELGs.  This reaffirms the observation that ELGs are not generally found in cluster environments.  As would be expected, the large peak at small separations is eliminated in the normal galaxy histogram.  At separations less than 3 Mpc the two distributions resemble each other more closely, but at larger separations they remain distinct.  A KS test reveals that the differences in these distributions are significant at the 98$\%$ confidence level.

The same trends may be discerned in the nearest distributions of the spring samples (Figure ~\ref{fig7}b).  In the analysis which includes all galaxies, the ratio of ELGs to normal galaxies at large separations (greater than 3 Mpc) is approximately 2:1.  This excess of large separations again decreases in the analysis from which cluster members have been removed, where the ratio is approximately 5:4 (see also Table 4).  Elimination of the cluster galaxies does not have as great of an impact on the spring normal galaxy distribution because there are fewer clusters in the spring region.  It has no effect on the ELG distribution.  KS tests shows that the ELG and normal galaxy distributions are different at the 99.0$\%$ confidence level for the samples in which all galaxies have been included, and 91.3$\%$ for the samples from which cluster members are removed.  The results of the nearest neighbor analyses for both fall and spring regions are consistent and show that the population of ELGs is less clustered than the population of normal galaxies.

Turning now to the nearest neighbor distributions for the dwarf/giant galaxy samples (Figure ~\ref{fig8}), it is important to bear in mind that the more pertinent analyses to examine are those in which cluster members have been excluded.  This is because the samples of ELGs are now being used to {\em trace} the population of dwarf galaxies and are no longer being investigated in and of themselves.  Thus, deconvolving selection effects caused by using the ELG population as the tracer is critical.  Plots for the samples in which all galaxies are included are shown only for the purposes of comparison.  In the fall analysis where cluster members have been removed (Figure ~\ref{fig8}a), there is an excess of dwarf nearest neighbor separations beyond 3 Mpc.  Whereas 26.6$\%$ (17/64) of the dwarfs have nearest neighbor separations greater than 3 Mpc, only 14.3$\%$ (33/231) of the normal galaxies are found in this regime.  A KS test applied to the fall dwarf and giant distributions in which cluster galaxies have been removed shows that the two samples differ at the 99.1$\%$ confidence level.  

In the spring dwarf/giant samples in which cluster galaxies have been removed (Figure ~\ref{fig8}b), there is also an excess of dwarfs at large separations.  27.3$\%$ (9/33) of the dwarfs compared to 15.0$\%$ (29/193) of the giants are found at separations greater than 3 Mpc.  There is also a substantial difference between the distributions at intermediate separations (2 Mpc to 6 Mpc).  The dwarf nearest neighbor histogram appears to be shifted to larger separations relative to the giant galaxy nearest neighbor histogram.  There is only one dwarf galaxy with a nearest neighbor separation less than 1 Mpc.  A KS test applied to these distributions indicates that the two samples differ at the 99.9$\%$ confidence level.  The results of the nearest neighbor analyses in the spring field are consistent with those of the fall field.   Both analyses show that intrinsically fainter, less luminous galaxies are less clustered than the more luminous giant galaxies.

\subsection{Correlation Functions}

To provide another statistical measure of clustering, correlation functions are computed for each group of data.  First, the spatial auto-correlation function is calculated for the control sample.  This statistic complements the nearest neighbor distributions in which control object-control object separations are measured.  Second, the cross-correlation function between the test sample galaxies and control-sample galaxies is determined.  This statistic complements the nearest neighbor distributions in which test object-control object separations are measured.  The differences in clustering between the two samples are quantified by the differences in the two correlation functions.

The method of computing correlation functions employed in this study is based on the discussion in Kirshner, Oemler, and Schechter (1979) and has been used by a number of authors, including Eder {\em et al.} (1989) and Salzer {\em et al.} (1990).  Operationally, the auto-correlation function is defined by the relation \[1 + \xi(r) = \sum_{i=1}^{n} \frac{\rho_{i}(r)}{\overline{\rho}}\] where $\rho_{i}(r)$ is the galaxian density at a distance {\em r} from the $i^{th}$ galaxy, $\overline{\rho}$ is the mean density of the entire sample volume, and {\em n} is the number of galaxies in the sample.  Following this prescription, for a given separation length {\em r}, spherical shells with radius {\em r} and width $\delta r$ (chosen to be 0.125 Mpc) are centered on each of the objects in the sample, and the galaxian densities for the shells $\rho_{i}(r)$ are computed and summed.  This sum is divided by $\overline{\rho}$ (determined from the luminosity function for the region) and yields $\xi(r)$ as given by the above expression.   The technique for computing the cross-correlation function is the same, except that the shells are centered on galaxies in the test sample and the shell densities are determined from the comparison sample.  In other words, $\rho(r)$ is the density of objects in the comparison sample at a distance {\em r} from an object in the test sample.  As in the nearest neighbor computations, where the computational comparison sample is used to search for nearest neighbors, the computational comparison sample is used here to compute shell densities.  All densities are corrected for decreased sampling with distance using the method described by Postman and Geller (1984).

Correlation functions for the galaxy samples are presented in Figure ~\ref{fig9} (ELG/normal galaxy sets) and Figure ~\ref{fig10} (dwarf/giant galaxy sets) where log $\xi(r)$ has been plotted as a function of log {\em r}.  The open circles represent the results from the auto-correlation analysis and the filled circles represent those from the cross-correlation analysis.  Error bars are estimated from the Poisson noise ($\sigma_{p}$) in each bin.  The solid lines are linear least-squares fits to the data and represent power-law functions of the form $\xi(r)=(r_{o}/r)^{\gamma}$ where the values of $\gamma$ and $r_{o}$ are indicated in each plot.  The fits are weighted such that the weight of each point is equal to $1/\sigma_{p}$ and the errors given for $r_{o}$ and $\gamma$ are associated with the uncertainties in the fit.

First, in examining the results of the correlation function analyses, it is reassuring to note that the auto-correlation function parameters (derived by fits to the open circles), calculated for the samples in which all galaxies are included, are generally consistent with the canonical values of 1.8 for $\gamma$ and 6.67 Mpc (5h$^{-1}$ Mpc with H$_{\circ}$=75 km s$^{-1}$ Mpc$^{-1}$) for $r_{o}$.  This is true for both the ELG/normal galaxy analysis and the dwarf/giant galaxy analysis. 

In the correlation function analyses for the ELG/normal galaxy samples, the correlation functions of the ELGs in all four plots tend to lie below that of the normal galaxies.  In previous studies (Salzer {\em et al} 1990, Rosenberg and Salzer 1994), the direct comparison of the auto- and cross-correlation functions are facilitated by the fact that the powers ($\gamma$) of the two functions are identical within their errors.  This allows for the calculation of a ratio of correlation amplitudes $\xi_{normal}/\xi_{ELG}$ that is independent of $r$.  However in this study, the values of $\gamma$ for the auto- and cross-correlation functions differ considerably, which makes a direct comparison of the correlation functions difficult.  Nevertheless, it is clear that the value of $\xi_{ELG}$ tends to be less than the value of $\xi_{normal}$, which implies that the ELGs are less clustered than the population of normal galaxies.  This is consistent with the results of the nearest neighbor tests and confirms the conclusion that the population of ELGs are less clustered than the population of normal galaxies.
 
To measure the difference in clustering, the ratio of the correlation function amplitudes, $\xi_{normal}(r)$ to $\xi_{ELG}(r)$, may be calculated at some fixed value of {\em r}.  From the nearest neighbor analyses, a separation of 3 Mpc appears to be the characteristic value at which the ELG and normal galaxy distributions diverge, so the ratio is evaluated at {\em r} = 3 Mpc.  The results are listed in Table ~\ref{tab4}.  In the samples where all galaxies are included, the corrlation function amplitude of the normal galaxies is approximately twice that of the ELGs.  The ratio drops to about 1.5 when cluster members are removed from the samples.  
 
The results of the dwarf/giant galaxy correlation function analysis also corroborate the results of the nearest neighbor tests of those samples.  In Figure 10 it is clear that the correlation functions of the dwarf galaxies tend to lie below that of the giant galaxies.  The disparity between the auto- and cross-correlation functions is particularly striking in the spring region.  For the analysis where cluster members have been removed, the correlation length of the giant galaxies is about 3 times larger than that of the dwarf galaxies; that is, the dwarfs in the spring region have much less clustering power than their giant galaxy neighbors.  This is most likely due to a combination of the relatively small number of galaxy clusters and prevalence of large voids in the spring region.

The ratios of correlation function amplitudes,  $\xi_{giant}$(3 Mpc) to $\xi_{dwarf}$(3 Mpc) are also shown in Table ~\ref{tab4}.  The values of the ratios for the UM Fall dwarf samples are comparable to those of the UM Fall ELGs.  In the UM spring region however, the ratios are significantly larger, again reflecting the characteristics of the galaxian distribution in the region.  Of course, the errors associated with the ratios in the spring region are large, so care must be taken when interpreting the magnitude of this result.

\section{Summary and Discussion}

This study uses emission-line galaxies (ELGs) drawn from the University of Michigan (UM) objective-prism survey to probe the spatial distribution of the dwarf galaxy population.  The UM survey provides two samples of ELGs which cover widely separated areas on the sky.  For each region a sub-sample of dwarfs (M$_{B} \geq $ -18.0) is analyzed relative to a sub-sample of higher luminosity  galaxies (M$_{B} < $ -18.0) from the Updated Zwicky Catalogue (UZC) which reside in the same volume of space.  The relative clustering properties of UM ELGs and UZC normal galaxies, regardless of luminosity, are also studied.  All quantitative analyses are carried out in the velocity range from 2000 km s$^{-1}$ to 8000 km s$^{-1}$.

\subsection{Galaxian Clustering and Activity}

In a previous paper which examined the relative spatial distribution of the UM ELGs, Salzer (1989) found that although ELGs follow the structures defined by normal galaxies, they tend to be less clustered and avoid regions of high galactic density.  Salzer (1989) also noted that some ELGs are found in regions otherwise devoid of normal galaxies.  With our new spectroscopic observations, inclusion of 18 additional ELGs from UM survey lists I-III, and a significantly more complete comparison sample of normal galaxies, our qualitative analysis of the cone diagrams of both fall and spring UM regions confirms and extends these conclusions.  Results from nearest neighbor and correlation function analyses also show that star-forming galaxies are preferentially found in lower density environments than non-active normal galaxies.  Nearest neighbor analyses reveal that the ELG samples have greater percentages of nearest neighbor separations at large values, while correlation function calculations indicate higher values of $\xi$ for the normal galaxy auto-correlation function as compared to the ELG-normal galaxy cross-correlation function.  The conclusion that ELGs are less clustered than the overall galaxy population may not be too surprising since ELGs tend to be of late morphological type.  However, even when the morphological make-up of the ELG and normal galaxy samples are homogenized by the removal of cluster galaxies, both tests still show a statistically significant difference between the clustering strength of the two populations.  Table ~\ref{tab4} summarizes the results from the nearest neighbor and correlation function analyses for all of the samples.

These results are consistent with other studies of the relative spatial distribution of ELGs.  Iovino et al. (1988), Rosenberg et al. (1994) and Loveday et al. (1999) have done correlation function analyses on ELGs and find a low clustering amplitude for this population.  Pustil'nik et al. (1994) show that about 15\% of their ELGs sample drawn from the Second Byurakan Survey and Case Low-Dispersion Northern Sky Survey are located within voids, indicating that active galaxies are more likely to inhabit regions of low galactic density than quiescent galaxies.  

A particularly puzzling issue which stems from these analyses concerns the cause of activity in ELGs.  A popular hypothesis is that galaxy-galaxy interactions are the triggers of wide-spread star formation.  While it is true that some ELGs are observed to be interacting with close companions (e.g., \cite{sal95}),  the result that ELGs tend to be more isolated than quiescent galaxies seems to suggest that interactions cannot be the primary cause of activity.  Of course, the current study has only cross-correlated ELGs with objects from surveys which are magnitude-limited at $\sim$15.5, so our conclusion is limited to interactions with bright galaxies ($M^{*} \pm$ 2).  Searches for objects in the vicinity of ELGs which have found a deficit of $ L > L^{*}$ galaxies within 1 Mpc (\cite{cam91,cam93,vil95,pus95,tel95}) are consistent with our nearest neighbor analyses, which show few ELG-normal galaxy pairs at small separations, and confirms that ELGs do not typically have giant companions.   

It has been proposed that the observed activity in ELGs may be caused by interactions with fainter dwarfs or HI clouds (\cite{bri90}).  However, recent work by Telles \& Maddox (1999) which has cross-correlated ELGs in the Spectrophotometric Catalogue of HII Galaxies (\cite{ter91}) - a sample which has selected the majority of its objects from the Tololo (\cite{smi76}) and the UM surveys - with faint field galaxies in the Automatic Plate Measuring Machine catalogues has concluded that the active galaxies in their sample are not preferentially associated with faint, low mass companions.  This leads to the general conclusion that interactions with galaxies, luminous or faint, are not the primary cause of the activity in ELGs.  It is not clear what may be inducing the star formation in these galaxies.  Although a large body of work is available that discusses triggering mechanisms involving tidal interactions (see for example \cite{ken98}), very little has been done to investigate the causes of global starburst activity in isolated galaxies.  Clearly, more effort must be invested in exploring alternate triggering mechanisms if further progress is to be made in understanding the formation and evolution of these galaxies.

\subsection{Galaxian Clustering and Luminosity}

As discussed in both $\S$1 and $\S$2, samples of ELGs do not suffer from the Malmquist effect as severely as magnitude-limited samples and have luminosity distributions which are more heavily weighted towards the faint end.  Moreover the redshifts of ELGs are easily determined, making this population not only an efficient probe of the dwarf galaxy population at distances well beyond the Local Supercluster, but also an effective tracer of the overall galaxian large-scale structure at distances where magnitude-limited samples become poorly sampled.

In our analyses of the relative spatial distribution of low- and high-luminosity galaxies, we find that faint dwarfs are not as strongly clustered as their giant counterparts.  There is an excess of dwarf-giant galaxy nearest neighbor separations of about 14\% at distances larger than 3 Mpc, and the ratio of the giant galaxy auto-correlation function to the dwarf-giant cross-correlation function is approximately 2.  Again, we have calculated these quantities for samples which we have attempted to homogenize with respect to morphology by removing all cluster galaxies.  

Our results are consistent with the scheme of ``natural bias" in galaxy formation models which has been discussed by many authors (e.g., \cite{whi87a,col89}).  These models speculate that galaxies have preferentially formed in the peaks of the initial density field.  For an initial field that is Gaussian, the larger but rarer 3$\sigma$-4$\sigma$ peaks would have been more clustered than the smaller, but more common 1$\sigma$-2$\sigma$ peaks.  Presumably, massive dark matter halos, which have emerged from the more extreme density fluctuations, have formed luminous giant galaxies, whereas low-mass halos, which have collapsed from the smaller peaks, have formed fainter, lower mass galaxies.  This leads to the prediction that high luminosity galaxies should be more clustered than the dwarf galaxies, which are less biased and better tracers of the overall mass distribution.  In particular, it is interesting to note the agreement of our calculations with the models of White et al. (1987a \& b) which predict differences between the correlation function amplitudes of luminous and dwarf galaxies of a factor of approximately 1.5 to 2.

Of course, it cannot be entirely ruled out that what is being detected is evidence that ELGs are less clustered than the population of normal galaxies, rather than evidence of biasing, since our dwarf samples are primarily composed of ELGs.  The excess number of nearest neighbor separations at large values and higher correlation function amplitudes are also found when statistics are calculated for the ELG/normal galaxy samples.  The differences in clustering strength for the dwarfs relative to the giants, however, are greater than the differences in that of the ELGs relative to the normal galaxies.  In the UM spring sample where cluster members have been eliminated, the ratio of dwarf nearest neighbor separations greater than 3 Mpc to that of the giant galaxies is larger when compared with the ratio of ELG nearest neighbor separations greater than 3 Mpc to that of the normal galaxies.  In the UM fall samples these  are approximately the same.  In all samples the ratio of correlation function amplitudes is larger for the dwarf/giant samples than for the ELG/normal galaxy samples.  Probably then, the differences in clustering strength between the emission-line selected dwarfs and the normal galaxy giants incorporate the effects of biasing {\em and} the phenomenon of diminished clustering with activity in galaxies.  Until the ELG activation mechanism and its possible dependence on environment is properly understood, however, the relative contribution of these two effects to the reduced clustering amplitudes of our dwarf ELG samples will not be conclusively known.

Further studies with larger and deeper samples of ELGs will allow us to more nearly deconvolve the dependence of clustering on activity and luminosity.  With larger samples, we will be able to compare the clustering of bursting and non-bursting galaxies in several luminosity bins.  In an upcoming paper (\cite{lee00}; Paper II), we analyze the relative spatial distribution of ELGs detected in the new, ultra-deep KPNO International Spectroscopic Survey (KISS) (\cite{sal99}).  KISS provides a sample of ELGs that is considerably deeper - up to two magnitudes fainter - than other existing surveys for active galaxies.  The area surveyed by KISS overlaps the Century Survey (\cite{gel97}) which covers the central 1$^{\circ}$ region of the first slice of the CFA Redshift Survey (\cite{del86}).  In Paper II, we undertake analyses which closely follow those performed in this study, and compare our results with other observational studies of the luminosity-density relation.  A discussion of the results of the current paper in the context of calculations by other groups will also be taken up there.

We gratefully acknowledge financial support for this research through an NSF Presidential Faculty Award to JJS (NSF-AST-9553020).  Early stages of this work were partially supported by a Cottrell College Science Award from the Research Corporation, as well as by research funds provided by Wesleyan University, to whom we are also grateful.  Thanks are due to the following staff members of Kitt Peak National Observatory for carrying out the queue observations on our behalf: Todd Boroson, Diane Harmer and Jim DeVeny.  We also thank Chris Impey and the anonymous referee for suggestions and comments that have improved the quality of this paper.

\clearpage
\begin{table}
\begin{center}
\tablenum{1}
\caption{New Spectroscopic Data}
\label{tab0}
\begin{tabular}{rrrrrrrrrrccrrrrrrr}
\tableline\tableline

\multicolumn{2}{c}{UM\ \#} & &$\alpha$ & \multicolumn{3}{c}{(J1950)} &$\delta$& && \ m$_{UM}$ & \ z & \multicolumn{3}{c}{EW$_{[O III]}$} &$\frac{[O III]}{H\beta}$ & \ $\frac{[O II]}{H\beta}$ & \ $\frac{[N II]}{H\alpha}$ & \ c$_{H\beta}$\\

(1) && \multicolumn{3}{c}{(2)} &\ \ & \multicolumn{3}{c}{(3)}&& (4) & (5) && (6) && (7) & (8) & (9) & (10)\\
\tableline \tableline
72 &&  0 & 47 & 20.5 &&  3 & 42 & 33 && 17.5 & 0.0714 &&  12.1 && 0.76 & 1.38 &&\\
 74 &&  0 & 48 &  8.2 &&  3 & 40 & 25 && 17.0 & 0.0729 &&  7.2 && 0.71 & 2.17 &&\\
113 &&  1 & 30 & 19.9 && 2 & 50 & 50 && 17.0 & 0.0434 &&  14.5 && 1.11 & 3.16 & 0.27 & 0.10\\
158 && 23 & 18 &  1.3 && -1 &  9 & 17 && 18.0 & 0.0145 && 206.2 && 4.09 & 2.42 & 0.04 & 0.00\\
206 &&  0 &  6 & 27.1 && -1 & 17 & 28 && 17.0 & 0.0389 &&  29.3 && 1.56 & 2.60 &&\\
235 &&  0 & 20 & 36.9 && -1 & 51 & 51 && 17.5 & 0.0645 &&   4.9 && 0.47 & 1.90 &&\\
242 &&  0 & 23 & 28.8 && -2 & 42 & 36 && 17.0 & 0.0562 &&  22.9 && 1.57 & 4.62 && 0.77\\
256 &&  0 & 30 & 41.1 && -0 &  4 & 28 && 16.5 & 0.0137 &&  12.3 && 1.59 & 4.33 & 0.16 & 0.10\\
257 &&  0 & 31 & 34.7 &&  0 & 17 &  9 && 17.0 & 0.0686 &&   2.6 && 0.53 & 2.23 &&\\
263 &&  0 & 36 & 30.3 && -2 & 15 & 41 && 17.0 & 0.0466 &&  68.5 && 3.79 & 2.96 && 1.38\\
280 &&  0 & 47 & 32.1 && -2 & 13 & 24 && 16.0 & 0.0127 &&  27.7 && 2.59 & 3.12 & 0.24 & 0.28\\
283 &&  0 & 49 & 15.6 &&  0 & 17 & 37 && 17.0 & 0.0153 && 122.4 && 3.14 & 3.64 & 0.07 & 0.00\\
286 &&  0 & 49 & 26.0 && -0 & 45 & 28 && 15.5 & 0.0055 && 141.7 && 3.03 & 3.03 & 0.07 & 0.00\\
289 &&  0 & 50 &  0.0 &&  0 &  3 & 36 && 17.0 & 0.0622 &&  86.4 && 2.73 & 2.97 &&\\
290 &&  0 & 50 & 24.1 &&  0 &  5 & 52 && 16.5 & 0.0340 &&  74.6 && 3.22 & 2.50 & 0.17 & 0.06\\
296 &&  0 & 56 & 30.1 &&  0 & 43 & 55 && 16.5 & 0.0177 &&  59.0 && 3.40 & 3.47 & 0.13 & 0.10\\
299 &&  0 & 59 & 57.6 &&  1 &  4 & 32 && 16.5 & 0.0164 &&  25.5 && 1.52 & 3.22 & 0.18 & 0.30\\
300 &&  1 &  0 & 15.2 && -1 & 44 & 31 && 17.0 & 0.0195 &&   9.1 && 1.37 & 4.68 & 0.17 & 0.41\\
\tableline \tableline
\end{tabular}
\end{center}
\end{table}

\clearpage
\begin{table}
\begin{center}
\tablenum{2}
\caption{Cluster Data \label{tab1}}
\vspace{.25in}
\begin{tabular}{cccccl}
\tableline\tableline
{\boldmath $\alpha$}& {\boldmath $\delta$}& {\bf Velocity}&{\bf Radius}& {\boldmath $\sigma_{vel}$}& {\bf Cluster}\tablenotemark{a}\\ 
{\bf(J1950)} & {\bf(J1950)} & {\bf(km s$^{-1}$)} & {\bf(degrees)} & {\bf(km s$^{-1}$)}& {\bf Name}\\ \tableline \tableline

\multicolumn{6}{c}{\bf UM Fall Region} \\
\tableline
-0$^{h}$ 41\fm6 & \plus8\fdg21 & 3400 & 0\fdg93 & 300\\
\plus0$^{h}$ 24\fm9 & -2\fdg03 & 4100 & 0\fdg74 & 190\\
\plus0$^{h}$ 26\fm2 & \plus2\fdg71 & 4300 & 0\fdg51 & 170\\
\plus0$^{h}$ 36\fm7 & \plus3\fdg01 & 4400 & 0\fdg27 & 300\\
\plus0$^{h}$ 36\fm8 & \plus0\fdg61 & 4300 & 0\fdg10 & 140\\
\plus0$^{h}$ 36\fm9 & \plus2\fdg84 & 5200 & 0\fdg85 & 110\\
\plus1$^{h}$ 08\fm9 & -0\fdg58 & 5500 & 1\fdg20 &170& SW 578\\
\plus1$^{h}$ 12\fm4 & \plus0\fdg03 & 13300 & 0\fdg70 &670& A 168\\
\plus1$^{h}$ 21\fm4 & \plus9\fdg19 & 2400 & 0\fdg59 & 200\\
\plus1$^{h}$ 21\fm6 & \plus1\fdg30 & 9700 & 1\fdg00&330& A 189\\
\plus1$^{h}$ 22\fm6 & \plus1\fdg37 & 5500 & 1\fdg20 &400& A 189B\\
\plus1$^{h}$ 23\fm4 & -1\fdg60 & 5400 & 1\fdg50& 500& A 194\\
\plus1$^{h}$ 53\fm7 & \plus5\fdg43 & 5500 & 0\fdg46 & 250\\
\plus2$^{h}$ 05\fm2 & \plus1\fdg84 & 7100 & 0\fdg20 & 150\\

\tableline \tableline
\multicolumn{6}{c}{\bf UM Spring Region} \\
\tableline

\plus11$^{h}$ 08\fm0 & \plus4\fdg74 & 8600 & 0\fdg48 & 230 &\\
\plus11$^{h}$ 09\fm2 & \plus3\fdg45 & 8900 & 0\fdg16 & 100 &\\
\plus11$^{h}$ 17\fm6 & \plus0\fdg01 & 7400 & 0\fdg47 & 140 &\\
\plus11$^{h}$ 47\fm0 & -3\fdg24 & 8200 & 0\fdg10 & 200 &\\
\plus12$^{h}$ 01\fm2 & \plus2\fdg18 & 5900 & 0\fdg99 & 450 & MKW 4\\

\tableline \tableline
\end{tabular}
\end{center}
\tablenotetext{a}{Names are provided only for clusters which have been identified in previous surveys.}
\end{table}

\clearpage
\begin{table}
\begin{center}
\tablenum{3}
\caption{Galaxy Samples \label{tab2}}
\vspace{.15in}
\begin{tabular}{cccc}
\tableline \tableline
&&&\raisebox{-.6ex}{\ipt Cluster} \\
\multicolumn{1}{c}{Test Population}&\multicolumn{2}{c}{Control Population}&{\ipt Galaxies}\\
{\ipt  2000 $\leq$ v [km s$^{-1}$] $\leq$ 8000} & {\ipt 2000 $\leq$ v [km s$^{-1}$] $\leq$ 8000} & {\ipt No Velocity Restrictions}\tablenotemark{a}&\raisebox{0.6ex}{\ipt Removed?}\\  
\tableline
{\ipt UM ELGs} & {\ipt Normal Galaxies in UM Area} & {\ipt Normal Galaxies in Extended Area}&{\foot No} \\
{\ipt UM ELGs} & {\ipt Normal Galaxies in UM Area} & {\ipt Normal Galaxies in Extended Area}&{\foot Yes} \\
{\ipt Dwarf Galaxies in UM Area} & {\ipt Giant Galaxies in UM Area} & {\ipt Giant Galaxies in Extended Area}&{\foot No}\\ 
{\ipt Dwarf Galaxies in UM Area} & {\ipt Giant Galaxies in UM Area} & {\ipt Giant Galaxies in Extended Area}&{\foot Yes}\\ 
\tableline \tableline
\end{tabular}
\end{center}
\tablenotetext{a}{Samples constructed for computational purposes.}

\vspace*{-.5in}

\tablenum{4}
\vspace{1in}
\caption{Summary of Nearest Neighbor and Correlation Function Analyses Results} 
\label{tab4}
\vspace{-1in}
\hspace{-2.9in}
\plotone{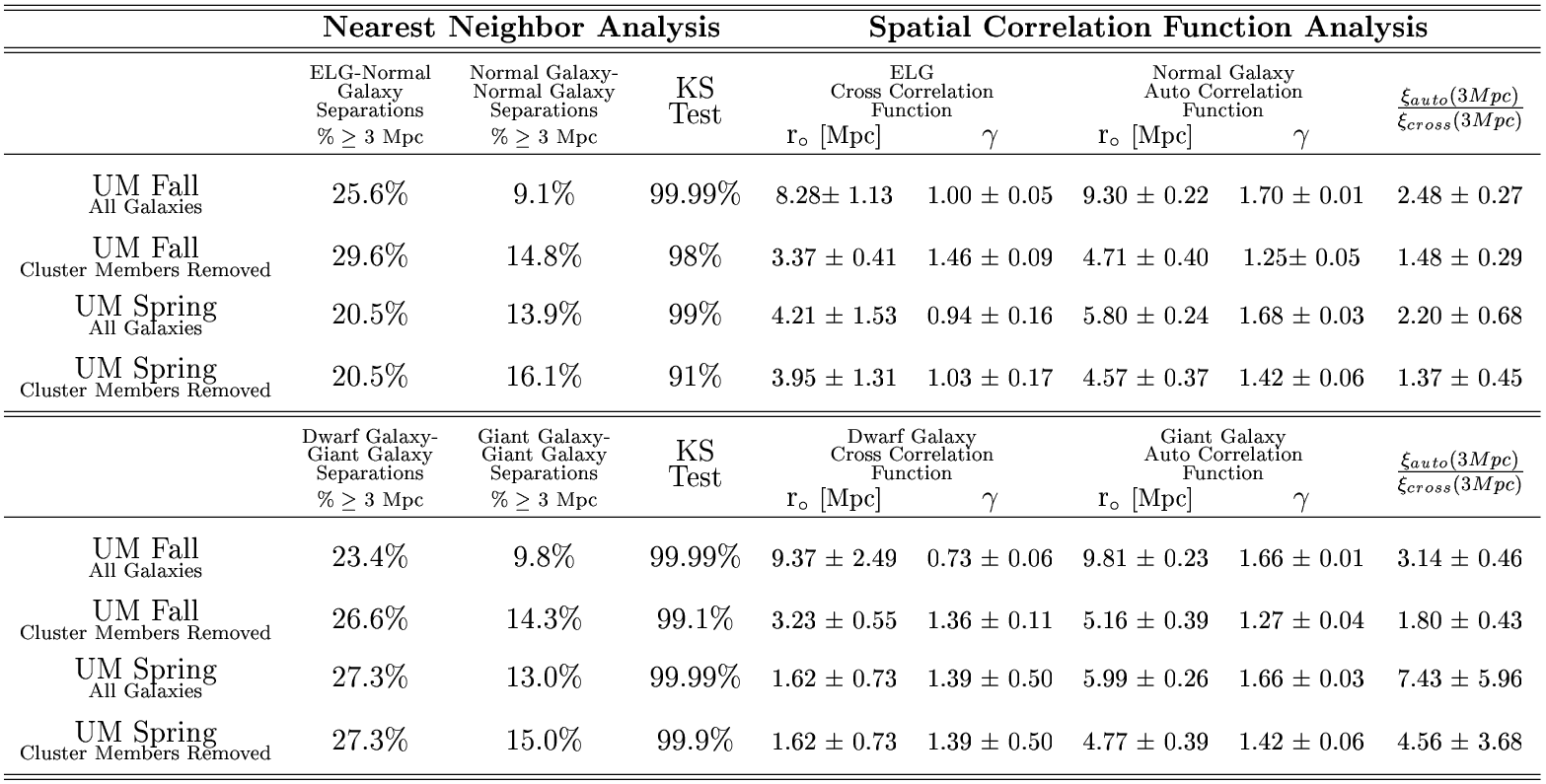}
\end{table}

\twocolumn

\onecolumn

\clearpage
\plotone{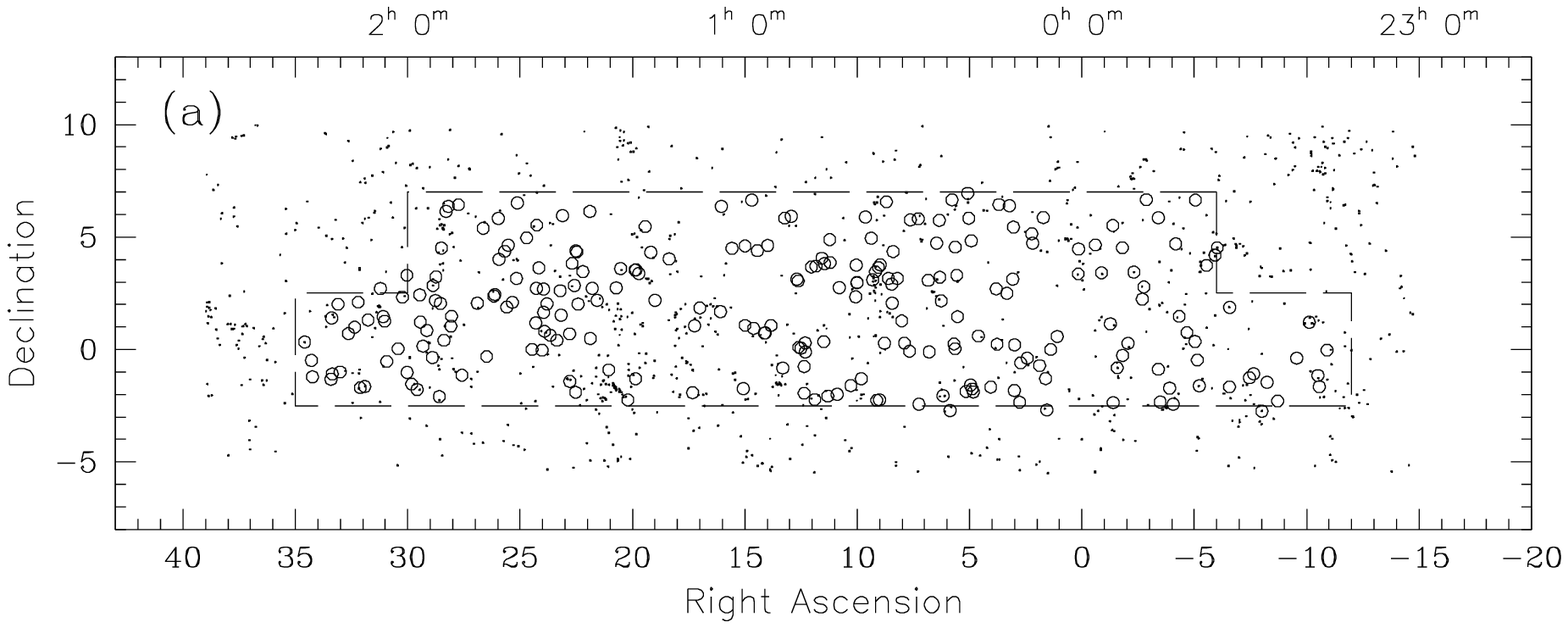}
\\
\plotone{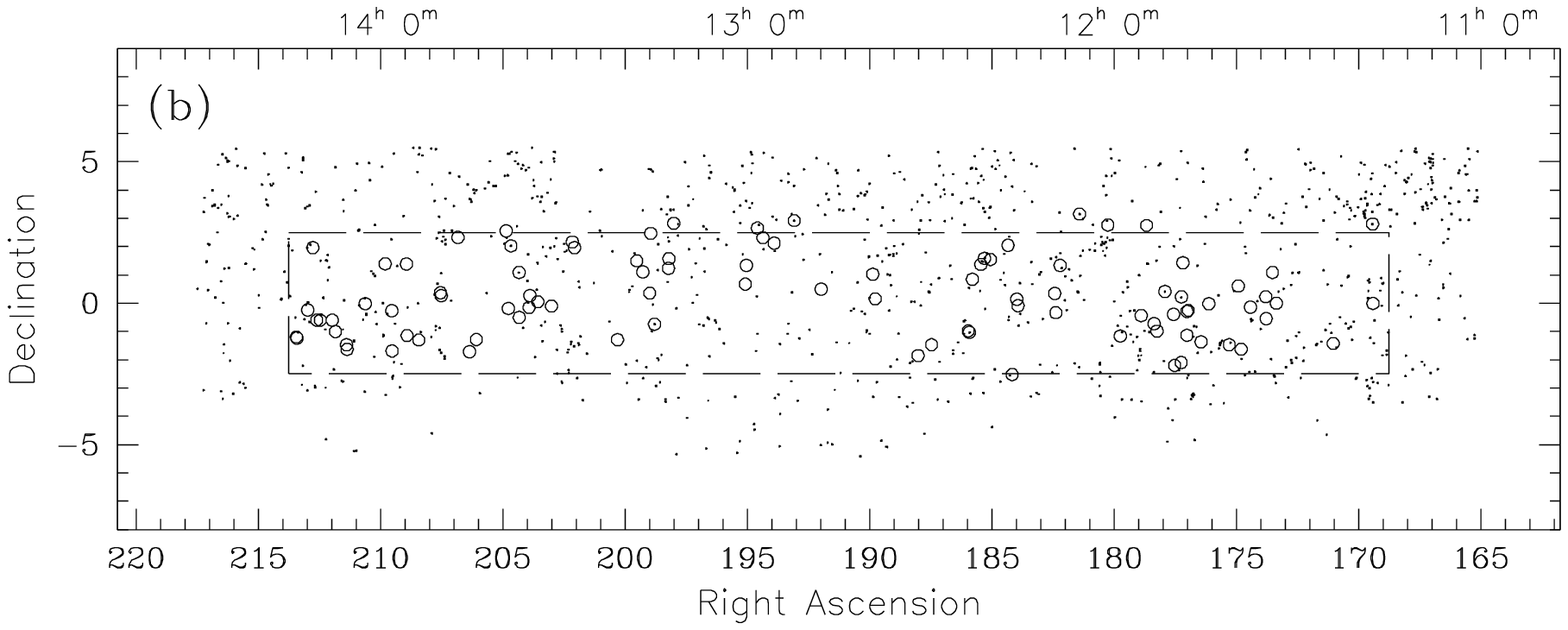}

\figcaption[fig1b.eps]{Sky positions of galaxies in the (a) fall and (b) spring UM survey areas. UM ELGs are plotted as open circles and UZC galaxies in the comparison sample are plotted as points.  The dashed lines indicate the boundaries of the UM survey. \label{fig1}}

\clearpage
\plotone{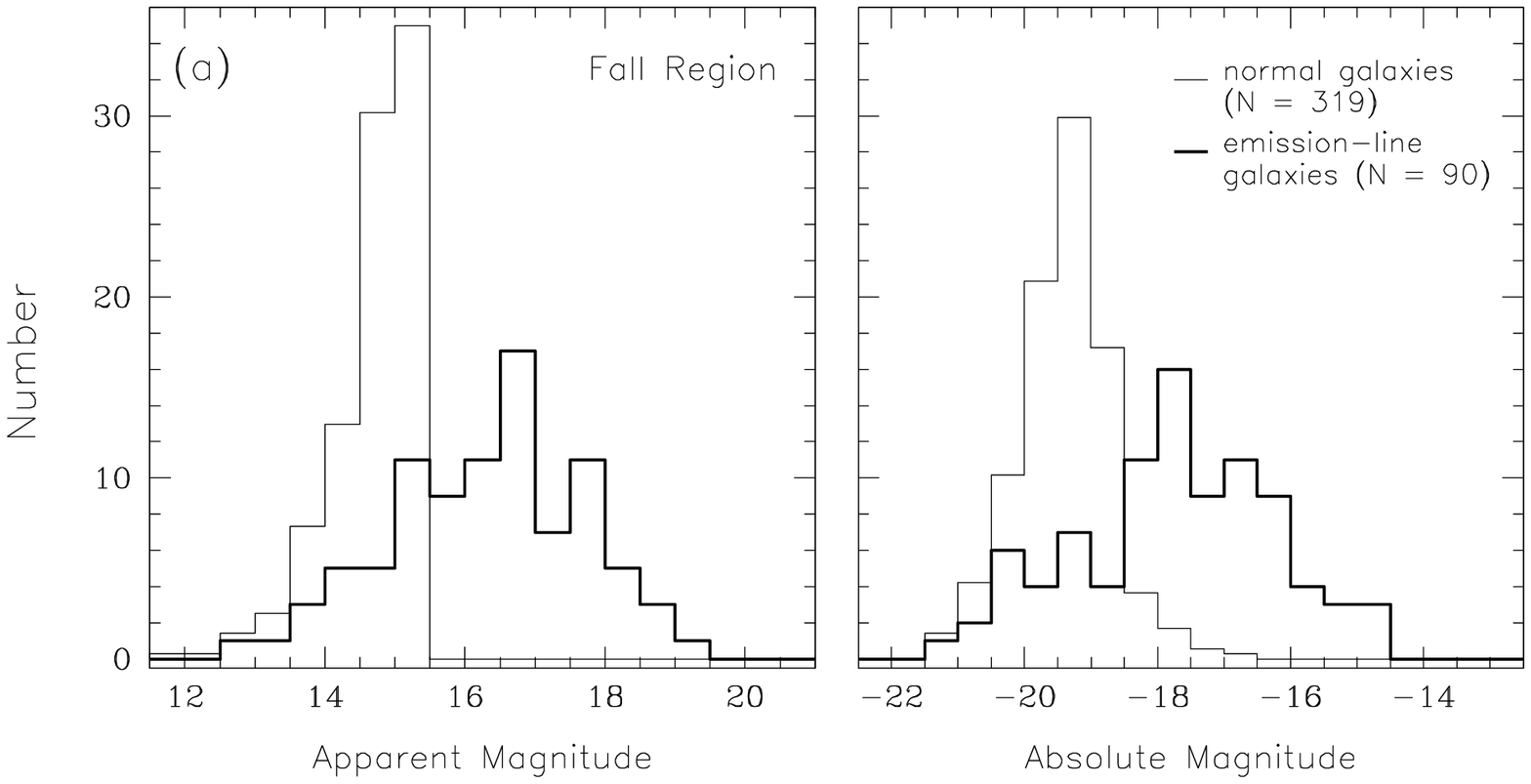}
\\
\plotone{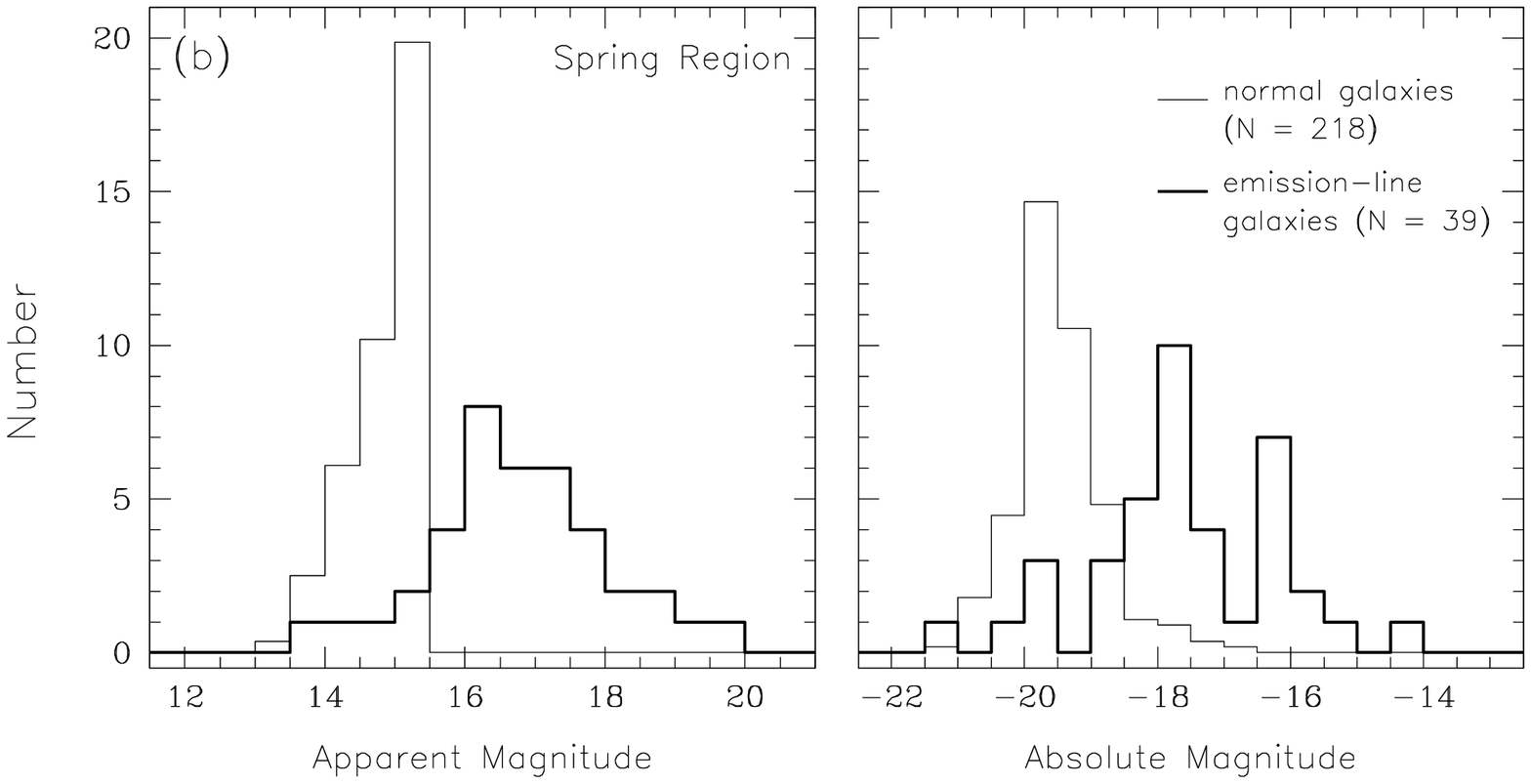}

\figcaption[fig2b.eps]{Apparent and absolute magnitude distributions for ELGs (heavy line) and UZC galaxies (thin line) within the velocity range from 2000 km s$^{-1}$ to 8000 km s$^{-1}$ in the (a) UM fall and (b) UM spring regions.  The areas under the UZC galaxy histograms are normalized to the areas under the ELG histograms. \label{fig2}}

\clearpage
\vspace*{-1.3in}
\plotone{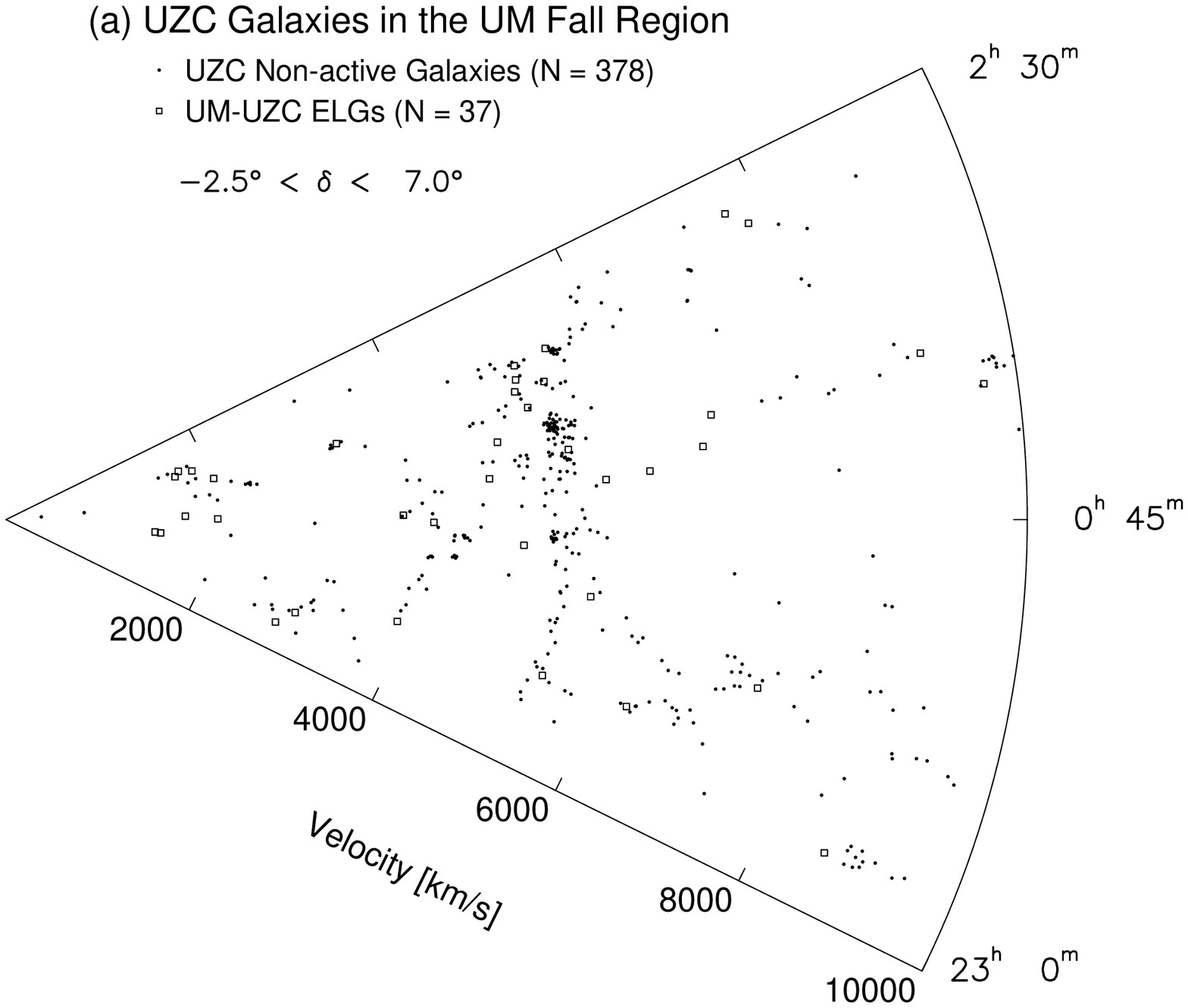}
\vspace*{-0.3in}
\nopagebreak
\vspace*{-.4in}
\plotone{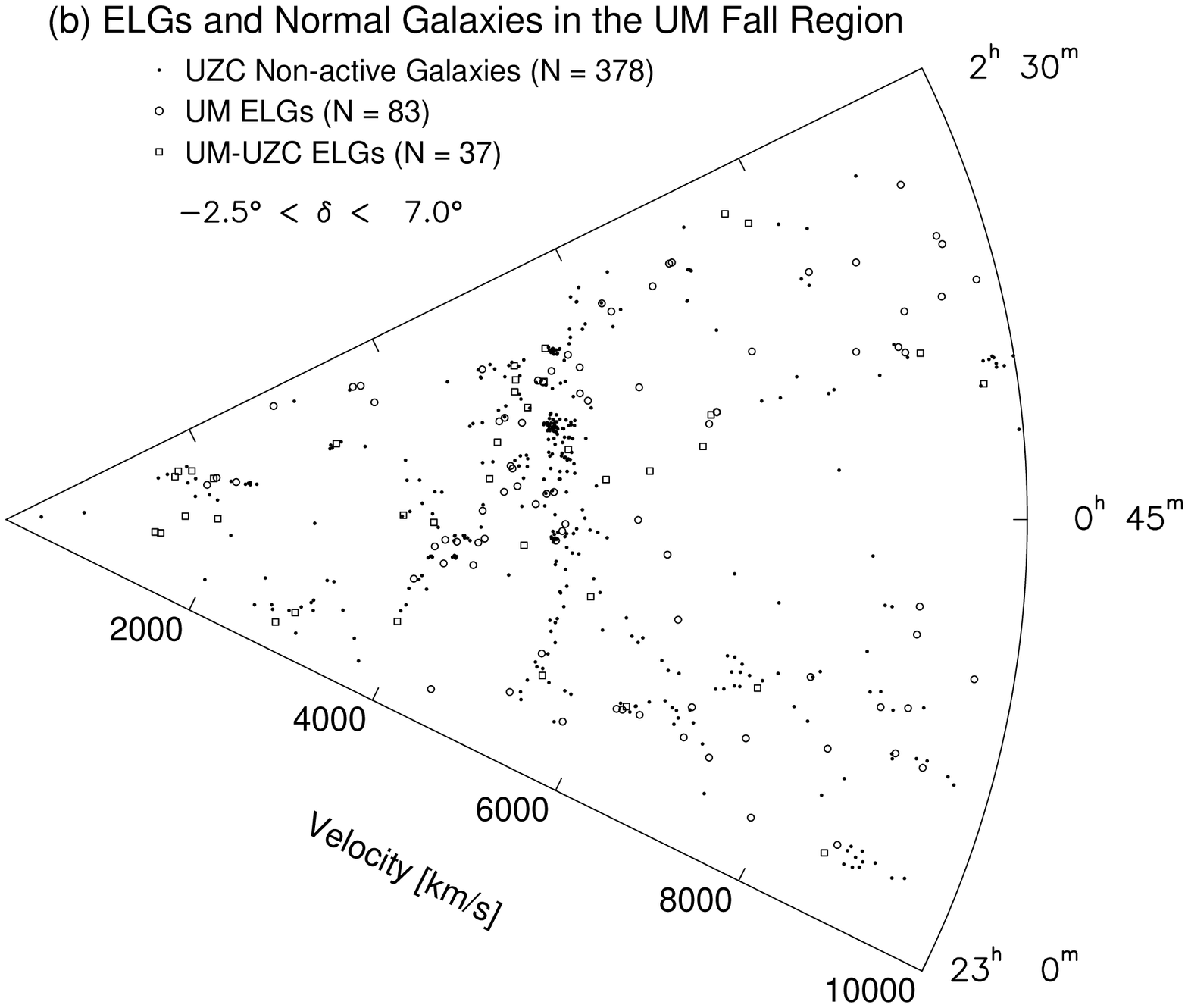}
\vspace*{-.6in}
\nopagebreak
\figcaption[fig3c.eps]{Cone diagrams for the region defined by lists I through IV of the UM survey.  Spatial distributions are illustrated for galaxies in (a) the UZC and (b) both the UM catalogue and the UZC.  In all diagrams, points represent normal (non-active) galaxies in the UZC, open circles represent ELGs in the UM catalogue and open squares represent ELGs which are detected in both surveys.  Note that the ELGs trace the same structures as the normal galaxies, but appear to be less tightly confined to them. \label{fig3}}

\clearpage
\vspace*{-1.2in}
\plotone{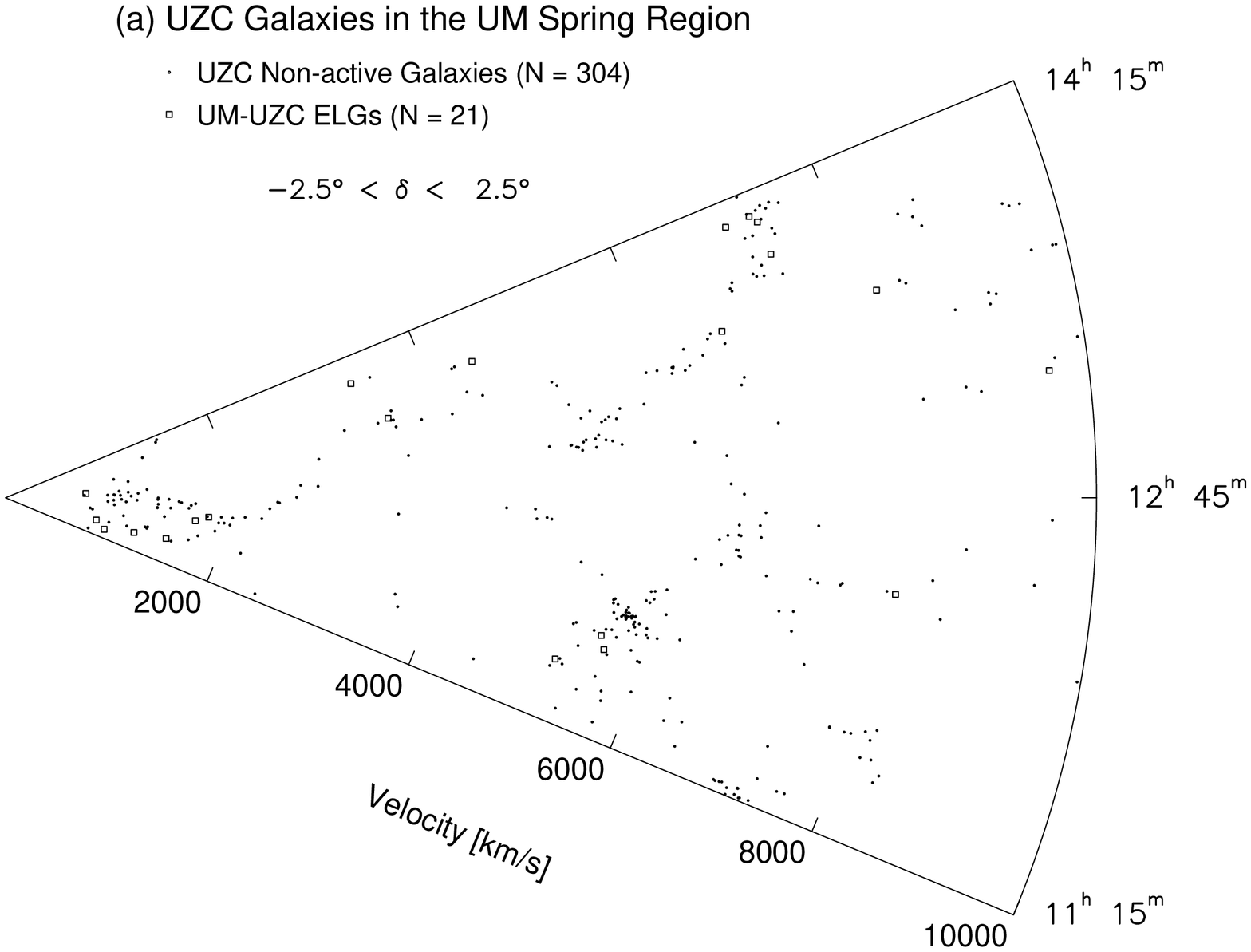}
\vspace*{-0.3in}
\nopagebreak
\vspace*{-.3in}
\plotone{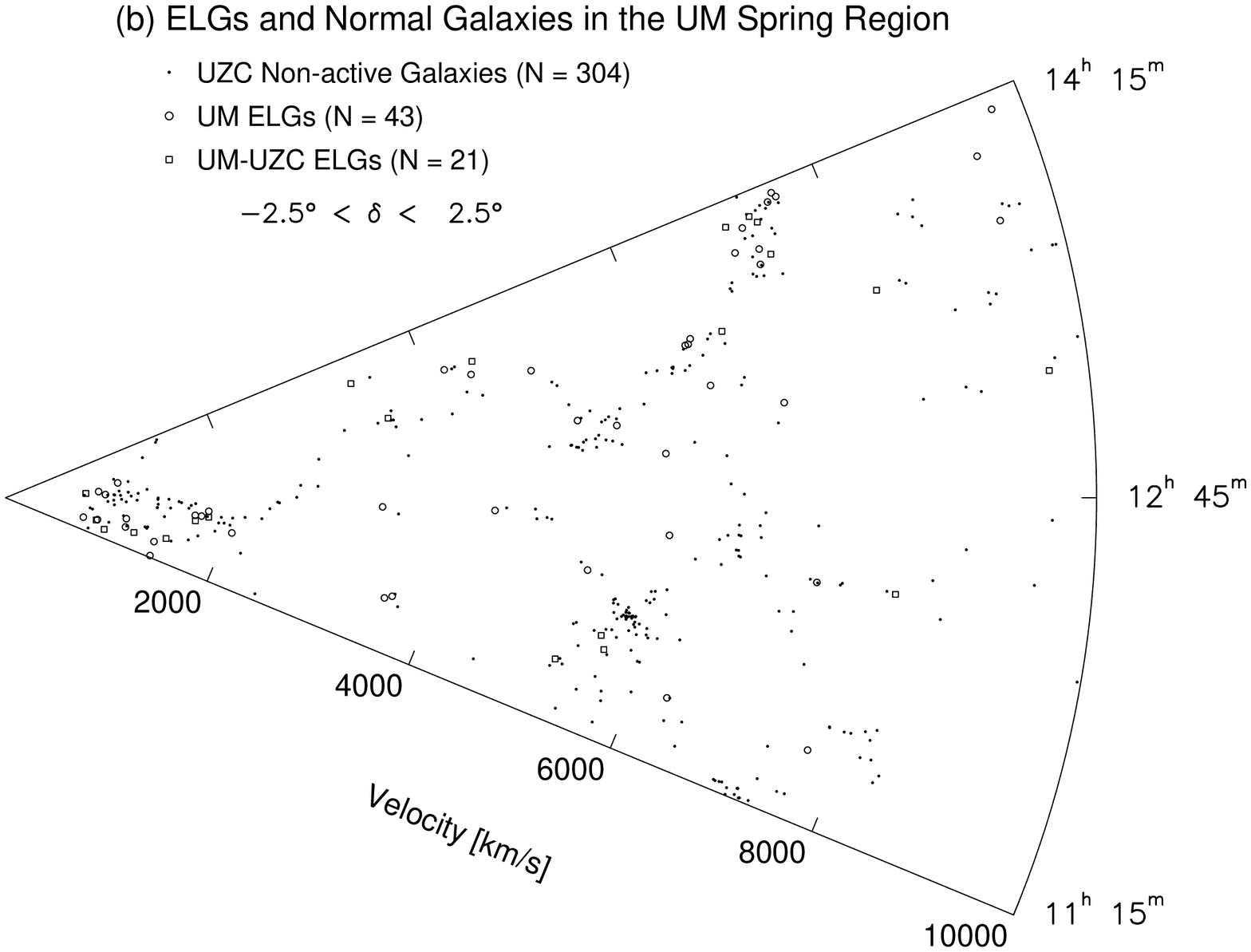}
\vspace*{-.6in}
\nopagebreak
\figcaption[fig4c.eps]{Cone diagrams for the region defined by list V of the UM survey.   Spatial distributions are illustrated for galaxies in (a) the UZC and (b) both the UM catalogue and the UZC.  In all diagrams, points represent normal (non-active) galaxies in the UZC, open circles represent ELGs in the UM catalogue and open squares represent ELGs which are detected in both surveys. \label{fig4}}

\clearpage
\vspace*{-1in}
\plotone{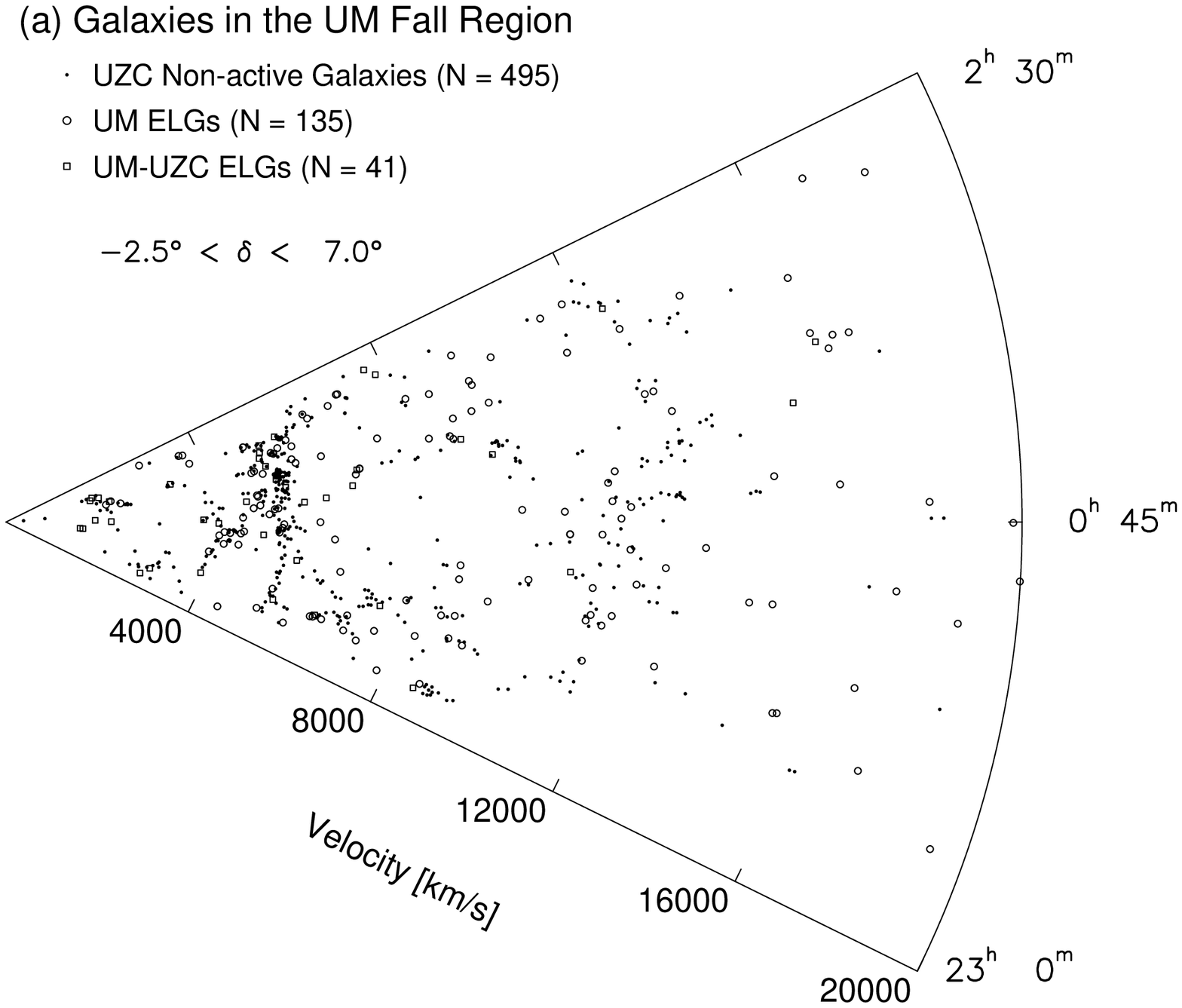}
\vspace*{-0.3in}
\nopagebreak
\vspace*{-.4in}
\plotone{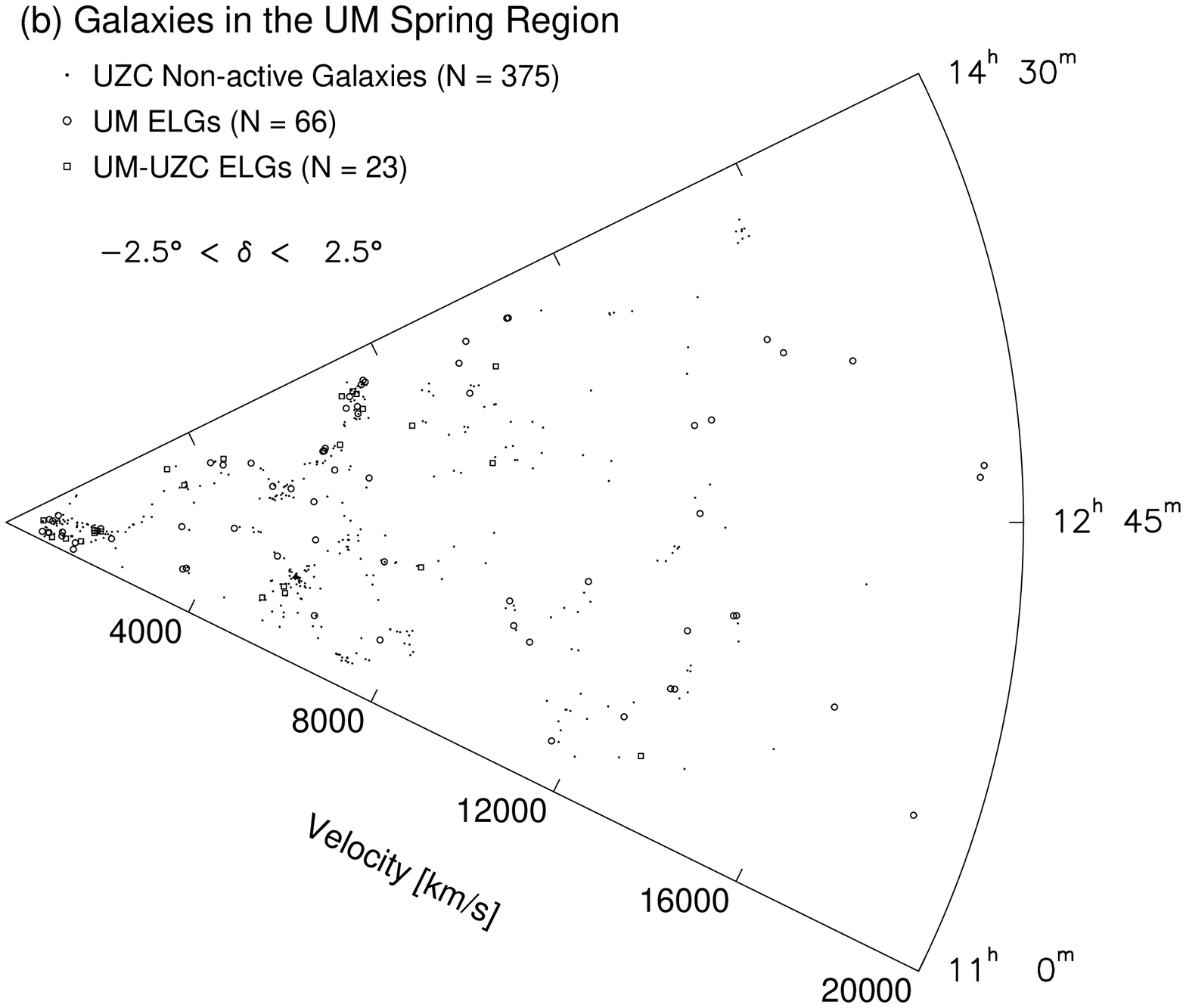}
\vspace*{-.6in}
\nopagebreak
\figcaption[fig5b.eps]{Cone diagrams for the (a) UM fall and (b) UM spring areas extended to 20,000 km s$^{-1}$. \label{fig5}}

\clearpage
\vspace*{.1in}
\plotone{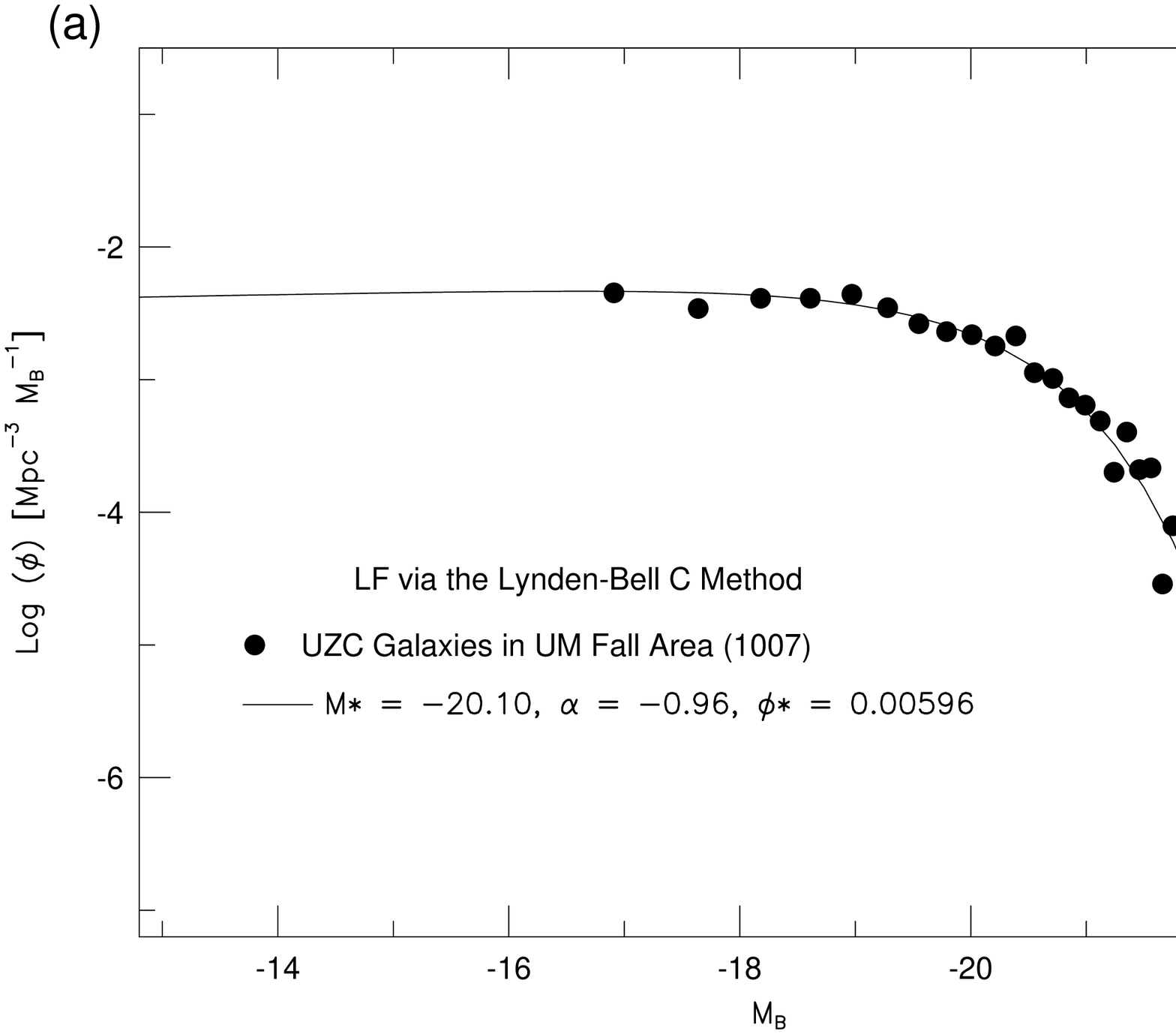}
\vspace*{0.2in}
\nopagebreak
\figcaption[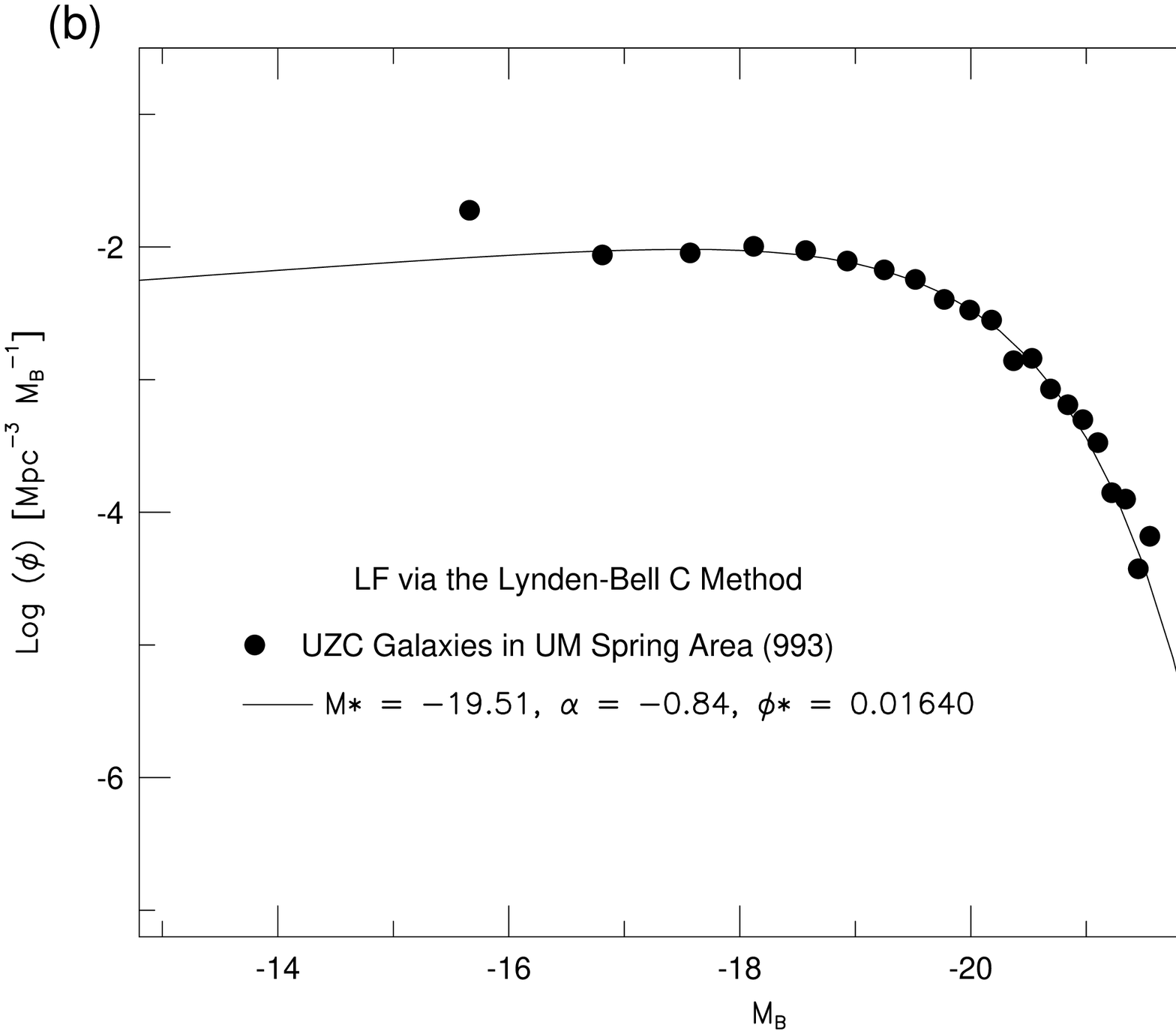]{Luminosity function calculated from galaxies in the UZC via the Lynden-Bell method for the (a) UM fall and (b) UM spring areas. \label{fig6}}
\nopagebreak
\vspace*{1.2in}
\nopagebreak
\plotone{fig6b.eps}

\clearpage
\vspace*{1in}
\plotone{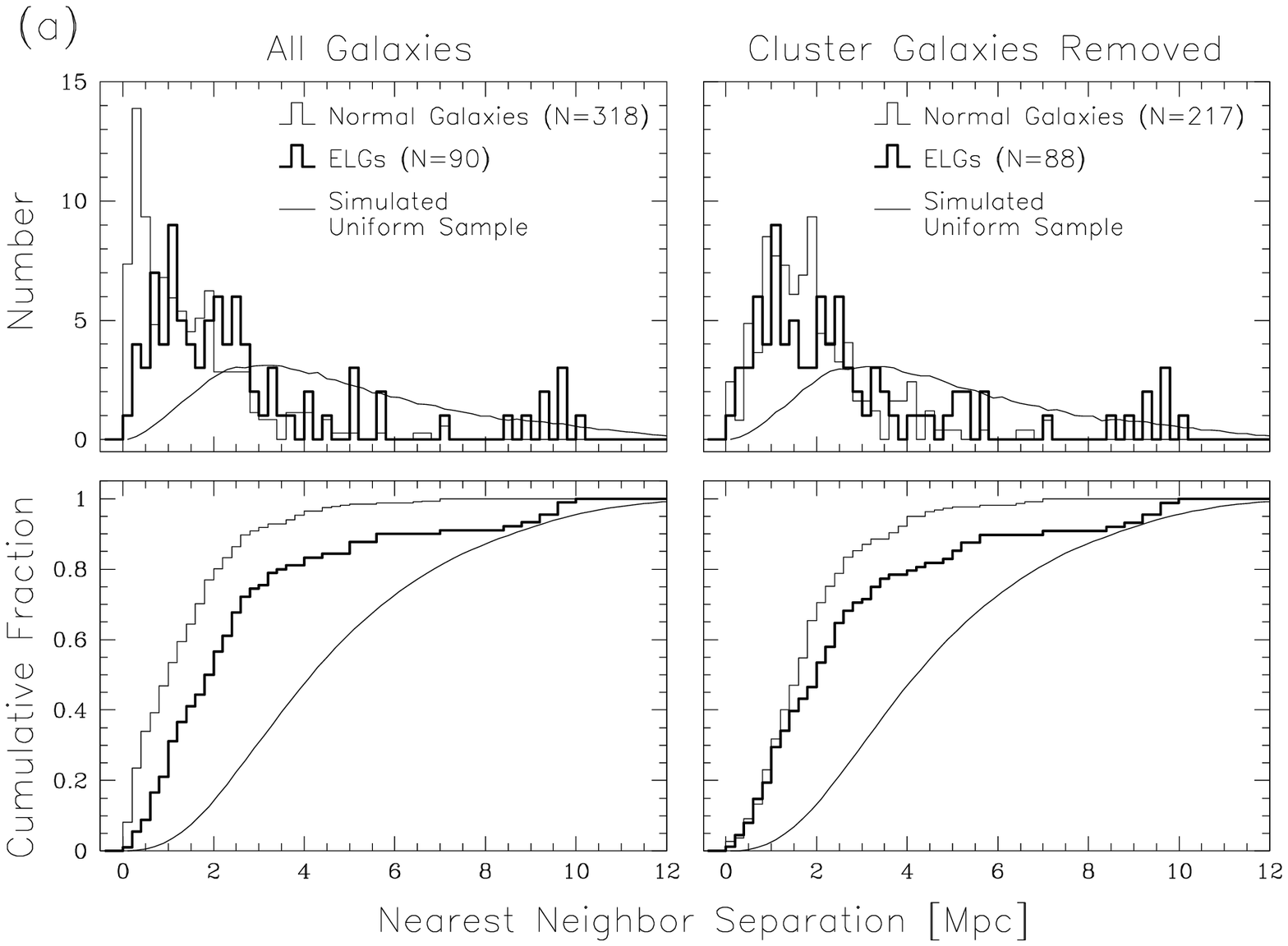}
\vspace*{-2in}
\figcaption[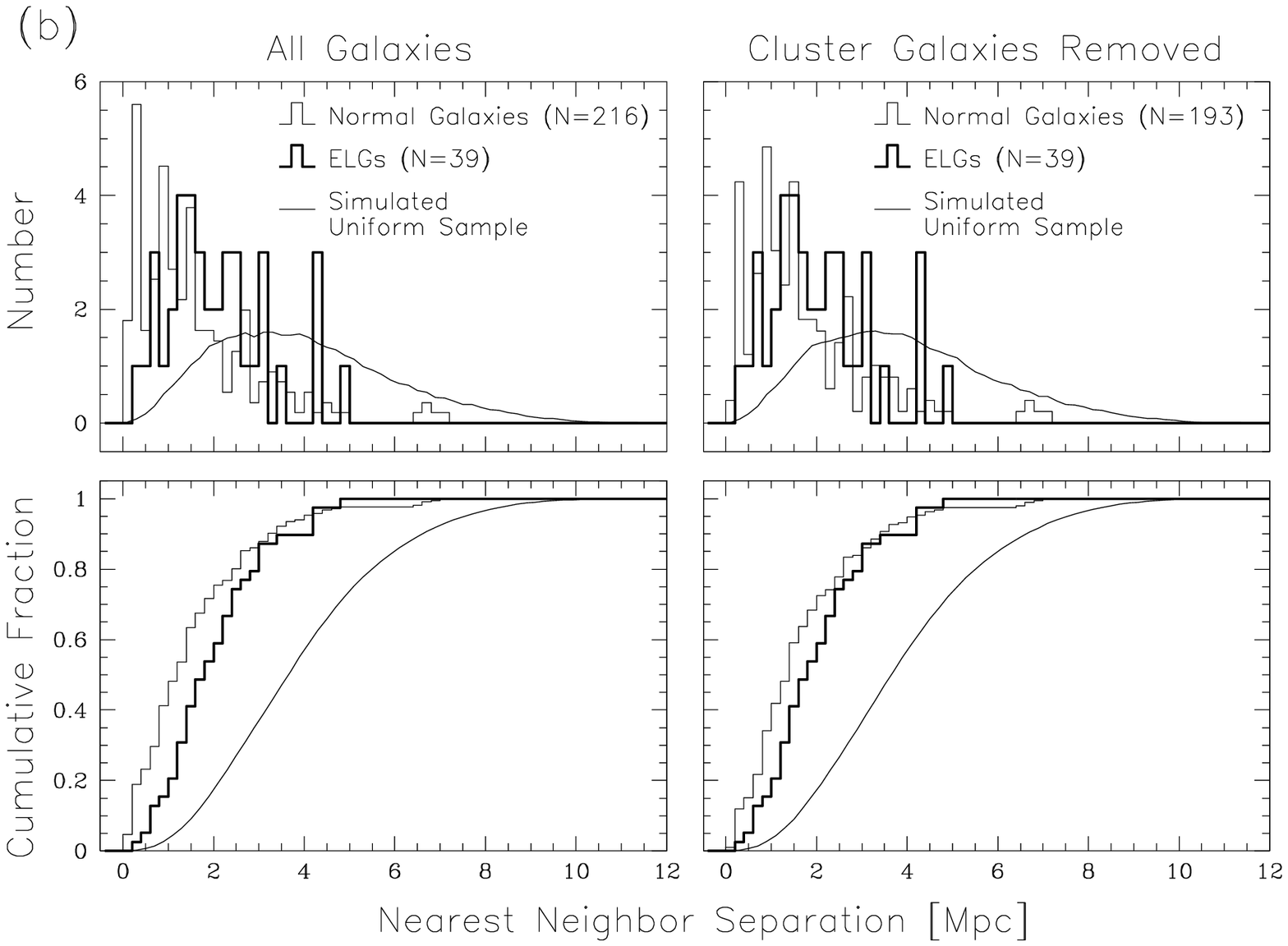]{Nearest neighbor distributions for the (a) fall (above) and (b) spring ELG/normal galaxy samples (next page). \label{fig7}}
\clearpage
\vspace*{1in}
\plotone{fig7b.eps}

\clearpage
\vspace*{1in}
\plotone{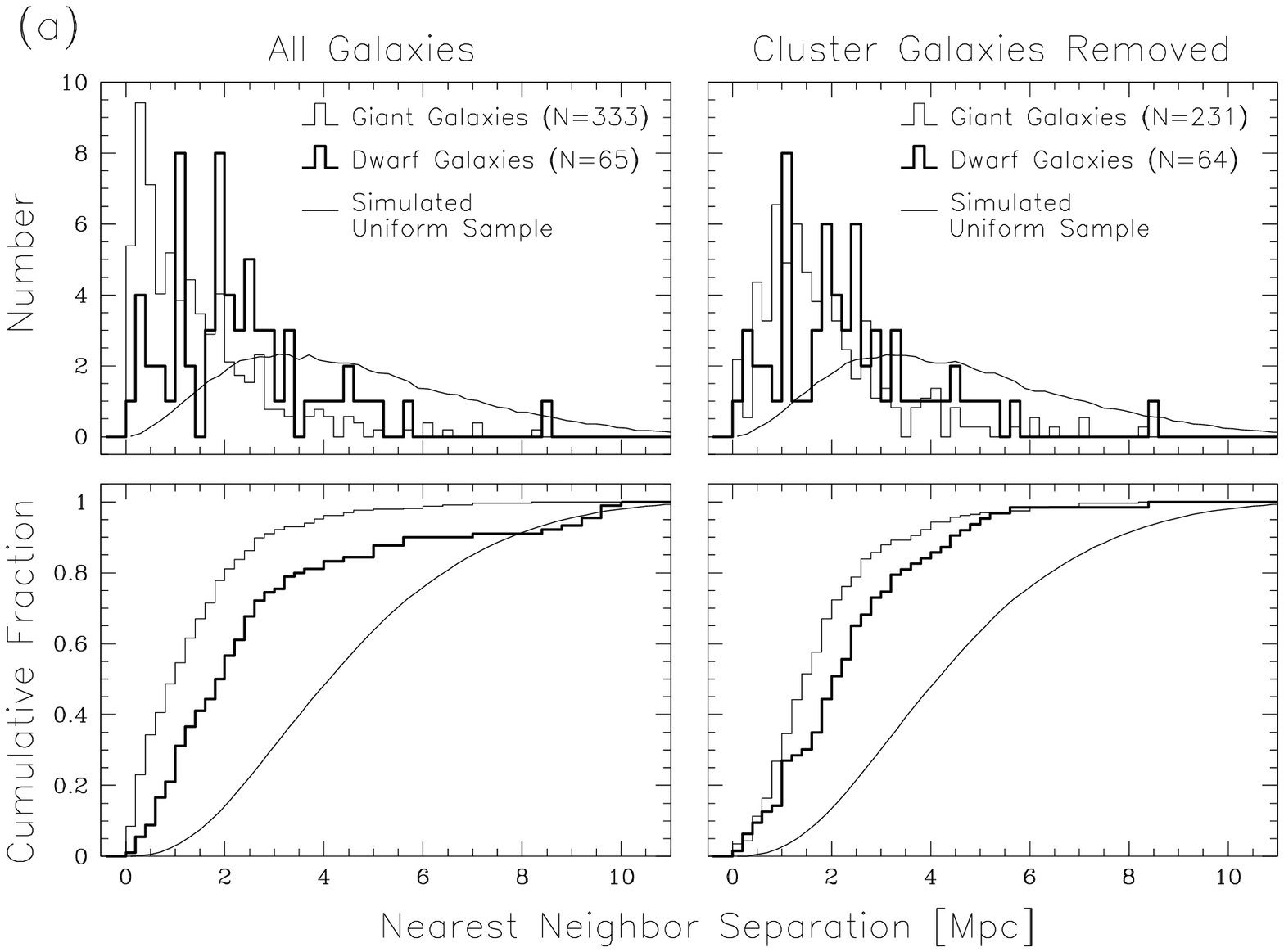}
\vspace*{-2in}
\figcaption[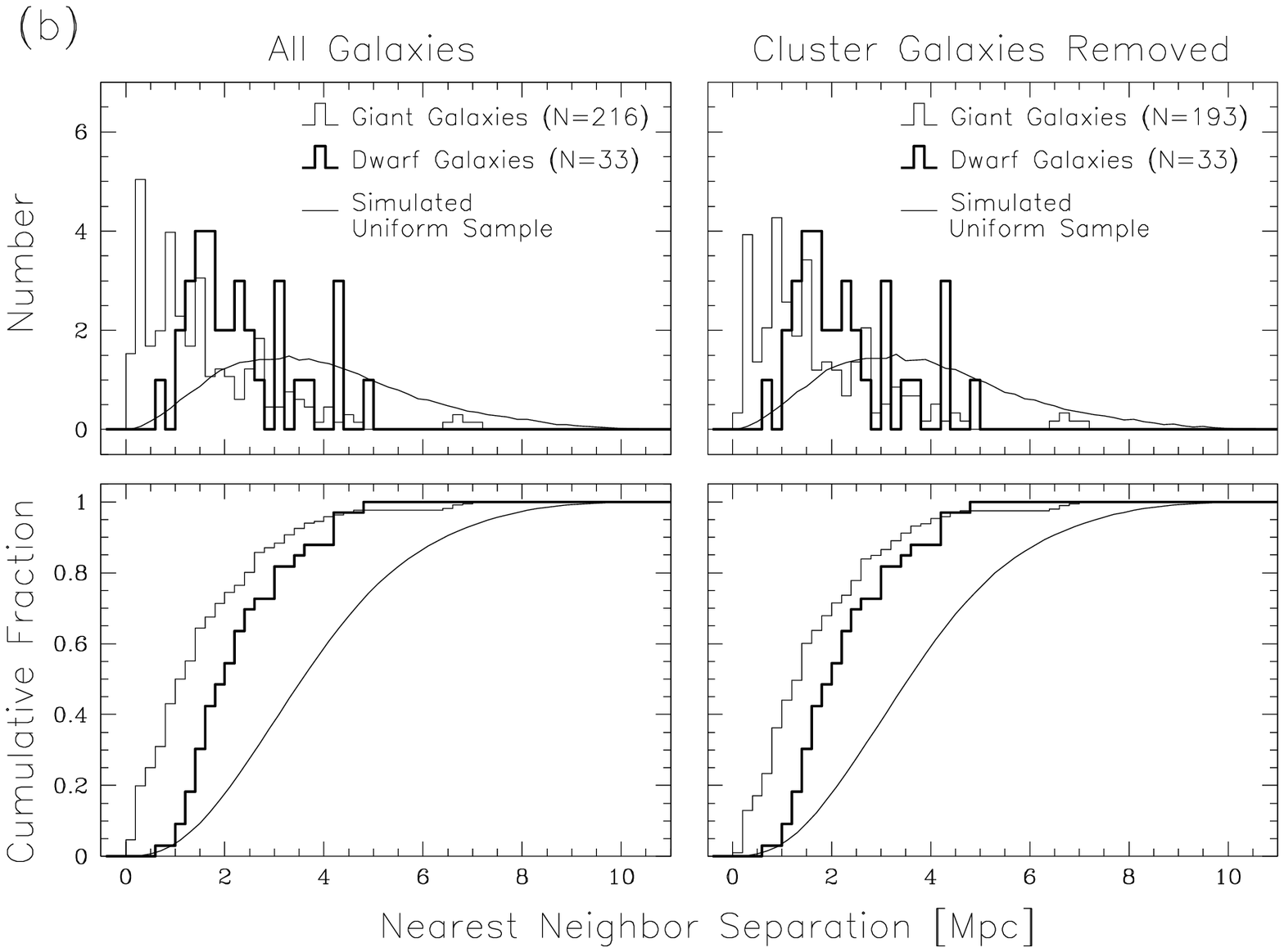]{Nearest neighbor distributions for the (a) fall (above) and (b) spring dwarf/giant galaxy samples (next page). \label{fig8}}
\clearpage
\vspace*{1in}
\plotone{fig8b.eps}

\clearpage
\vspace*{1in}
\plotone{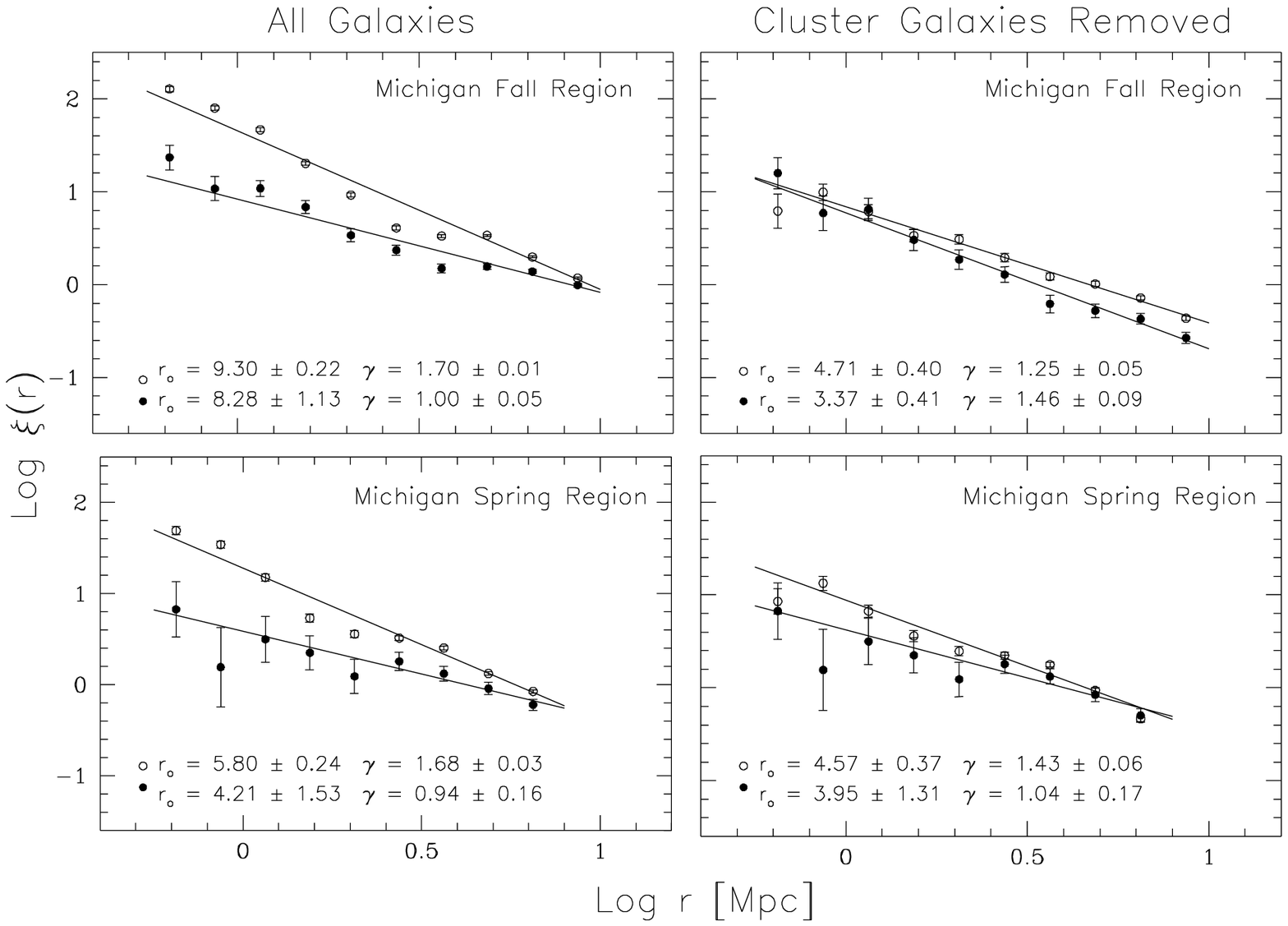}
\vspace*{-2in}
\figcaption[fig9.eps]{Correlation functions of the ELG/normal galaxy samples.  Auto-correlation function data points are indicated by open circles and cross-correlation data points by filled circles.  In all diagrams, the data are fit with power-law functions of the form $\xi(r)=(r_{o}/r)^{\gamma}$ where $\gamma$ and $r_{o}$ are specified in each plot.  Error bars are estimated from the Poisson noise in each bin. \label{fig9}}

\clearpage
\vspace*{1in}
\plotone{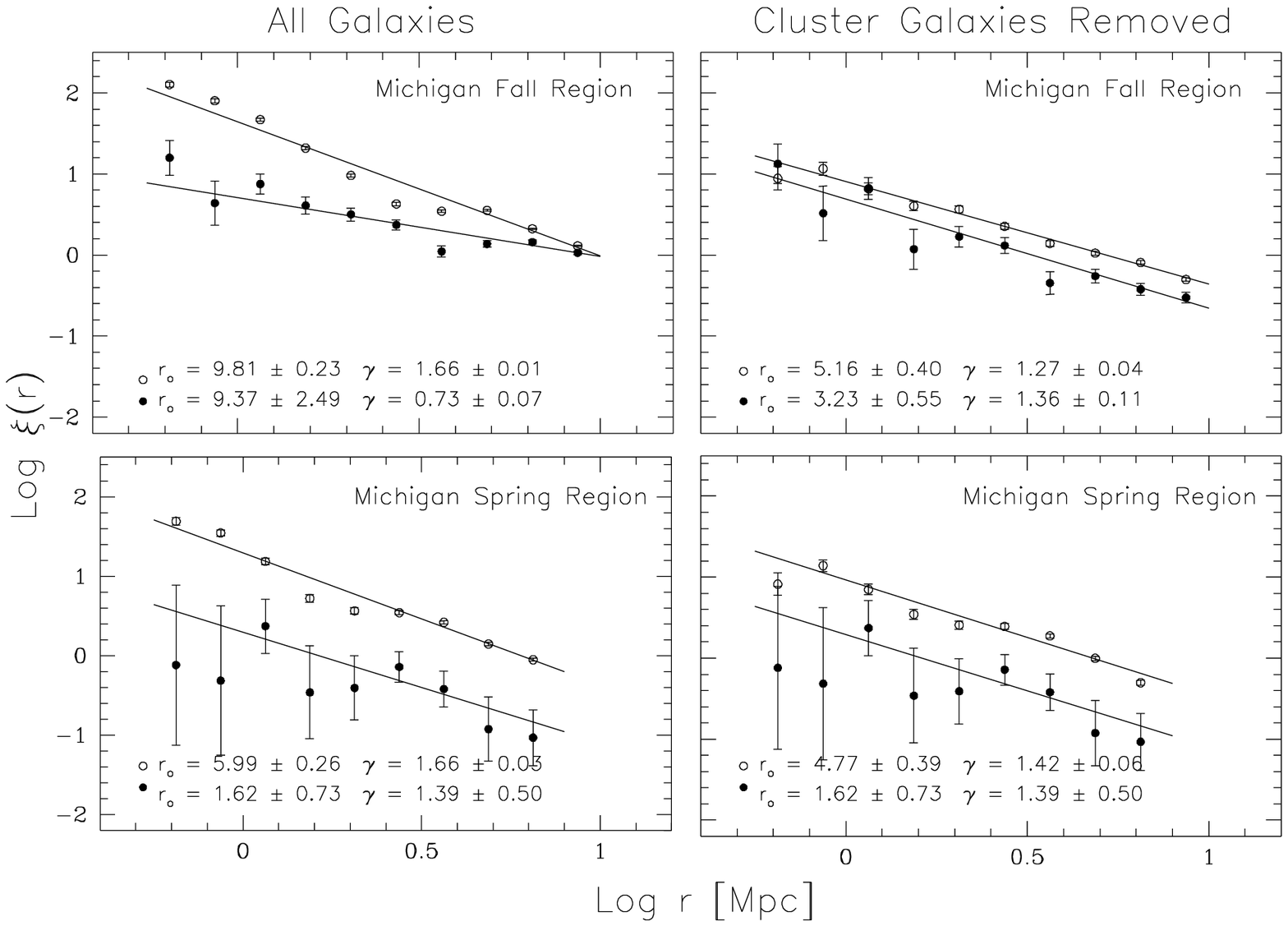}
\vspace*{-2in}
\figcaption[fig10.eps]{Correlation functions of the dwarf/giant galaxy samples.  As in Figure ~\ref{fig9}, auto-correlation function data points are indicated by open circles and cross-correlation data points by filled circles.  Error bars are estimated from the Poisson noise in each bin.\label{fig10}}

\end{document}